\documentclass{JHEP3}
\usepackage{epsfig,multicol,bbm}
\usepackage{makeidx}
\usepackage{amsmath}
\usepackage{amsfonts}
\usepackage{amssymb}
\usepackage{graphicx}
\usepackage{accents}
\input{tcilatex}

\newcommand\fverb{\setbox\fverbbox=\hbox\bgroup\verb}
\newcommand\fverbdo{\egroup\medskip\noindent%
            \fbox{\unhbox\fverbbox}\ }
\newcommand\fverbit{\egroup\item[\fbox{\unhbox\fverbbox}]}
\newbox\fverbbox

\title{\textbf{ \textbf{Anomalous Quantum Hall Effect of
4D Graphene in Background Fields}}}

\author{L.B Drissi $^{1,3}$, H. Mhamdi $^{1,2}$, E.H Saidi $^{1,2,4}$\\
  \small {1.INANOTECH, Institute of Nanomaterials and Nanotechnology (MAScIR), Morocco,} \\
 \small {2.Lab HEP-Modeling and Simulation, FSR, Univ Mohammed V-Agdal, Rabat, Morocco,}\\
  \small {3.International Centre for Theoretical Physics, ICTP, Trieste, Italy,} \\
 \small {4.Centre of Physics and Mathematics, CPM-CNESTEN, Rabat,
Morocco,} \\
   E-mails: \email{ldrissi@ictp.it}, \email{h-saidi@fsr.ac.ma}}

 \preprint{LPHE-11-05/ CPM-11-05}

\abstract{Bori\c{c}i-Creutz (\emph{BC}) model describing the
dynamics of light quarks in lattice \emph{QCD} has been shown to be
intimately linked to the four dimensional extension of 2D graphene
refereed below to as four dimensional graphene (\emph{4D-
graphene}). Borrowing ideas from the field theory description of the
usual \emph{2D} graphene, we study in this paper the anomalous
quantum Hall effect (AQHE) of the \emph{BC} fermions in presence of
a constant background field strength $\mathcal{F}_{\mu \nu }$ with a
special focuss on the case $\mathcal{F}_{\mu \nu
}=\mathcal{B}\varepsilon
_{\mu \nu 34}+\mathcal{E}\varepsilon _{12\mu \nu }$ with $\mathcal{B}$ and $%
\mathcal{E}$ two real\ moduli and $\det \mathcal{F}_{\mu \nu }=\mathcal{B}%
^{2}\mathcal{\times E}^{2}$. First, we revisit the anomalous
\emph{2D} graphene by using \emph{QFT} method. Then, we consider the
AQHE of \emph{BC}
fermions for both regular $\det \mathcal{F}_{\mu \nu }\neq 0$ and singular $%
\det F_{\mu \nu }=0$ cases. We show, amongst others, that the exact
solutions of the \emph{BC} fermions coupled to constant
$\mathcal{F}_{\mu \nu }$ have a 5D interpretation; and the filling
factor $\nu _{BC}$ of the
\emph{BC}\ model coupled to constant $\mathcal{F}_{\mu \nu }$ is given by $24%
\frac{\left( 2N+1\right) \left( 2M+1\right) }{2}$ with $N,$ $M$
positive integers. Others features, such as $\mathcal{F}_{\mu \nu
}^{QCD}\neq 0$ \textrm{and the extension of the obtained results to
the lattice fermions like Karsten-Wilzeck (KW) fermions and naive
ones}, are also discussed.}


\begin{document}


\section{Introduction}

\begin{figure}[tbph]
\begin{center}
\hspace{0cm} \includegraphics[width=8cm]{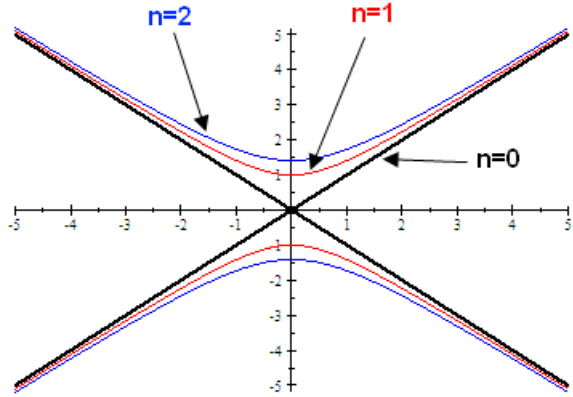}
\end{center}
\par
\vspace{-1cm}
\caption{{\protect\small t}he first levels of the energy spectrum $%
E_{n}^{\pm }\left( \protect\varpi ,k\right) =\pm \hbar
\protect\varpi \protect\sqrt{n+\frac{k^{2}}{\protect\varpi ^{2}}}$
of \emph{BC} fermions in
the background field{\protect\small \ }$\mathcal{F}_{\protect\mu \protect\nu %
}=\mathcal{B}\protect\varepsilon _{\protect\mu \protect\nu 34}$ with
$\det \mathcal{F}_{\protect\mu \protect\nu }=0$. For $n=0$, the
conducting and valence bands touch at the Dirac point $k=0$.}
\label{QH}
\end{figure}

Few years ago M. Creutz made a remarkable observation about links
between the physics of \emph{2D} graphene and four dimensional
lattice
chromodynamics (QCD) \textrm{\cite{1A,1B}}. The two Dirac valleys of \emph{%
2D }graphene \textrm{\cite{1C}-\cite{1E}} have an analogue in the
\emph{4D} extension of Creutz; and are interpreted as the up and
down quarks of the physics of QCD{\normalsize \
}\textrm{\cite{2A,2B,2BA,2BB,2BC}}. This correspondence opened a
window on applying constructions of \emph{2D} graphene modelings to
\emph{4D} lattice QCD formulated on the hyperdiamond lattice
\textrm{\cite{2A}-\cite{2F}}. \newline In this paper, we contribute
to this matter by looking for the
generalization of some specific properties of graphene to the case of \emph{4%
}D lattice QCD with a special focus on Bori\c{c}i-Creutz (BC)
fermions \textrm{\cite{3A}-\cite{3F}}; in particular the aspect
regarding the anomalous quantum Hall effect (AQHE)
\textrm{\cite{4A}-\cite{4J}}. Recall that QHE in higher dimensions
has been first considered in \cite{4F} for the
case of the 4-sphere and has been further developed in{\normalsize \ \textrm{%
\cite{4G,4H,4I}} }and refs therein for other higher D- manifolds.
These
models are non relativistic systems generalizing the well known 2D Hall fluid%
{\normalsize \ }\textrm{\cite{4IA,4IB}} where, due to disorder, the
conductivity has plateaus resulting from the non uniformity of the
spatial potential. Recall also that the quarks $u_{\alpha }\left(
\mathbf{x}\right) $ and $d_{\alpha }\left( \mathbf{x}\right) $ are
light particles that play a central role in lattice QCD simulations;
they are described by 4D Dirac spinors $\Psi _{\alpha }\left(
\mathbf{x}\right) $ with integral color quantum numbers as required
by the $SU_{c}\left( 3\right) $ non abelian gauge symmetry; but
fractional electromagnetic charges respectively\ equal to
$Q_{u}=\frac{2e}{3}$ and $Q_{d}=-\frac{e}{3}$. Interactions between
quarks are mediated by gauge fields; in particular by $A_{\mu }^{G}=$ $%
A_{\mu }^{_{em}}Q_{em}$ $+$ $\sum_{I}A_{\mu }^{I}\mathcal{T}_{I}$
valued in the adjoint representation of the gauge symmetry
$G=U_{em}\left( 1\right) \times SU_{c}\left( 3\right) $ with the
\emph{8} hermitian $3\times 3$
matrices $\mathcal{T}_{I}$ standing for the basis generators of $%
SU_{c}\left( 3\right) $. In the QCD regime \textrm{\cite{1G,2G,3G}}
where the $SU_{c}\left( 3\right) \times U_{em}\left( 1\right) $
gauge interactions can be approximated by \emph{constant} background
fields $\mathcal{F}_{\mu \nu }^{G}=(\mathcal{F}_{\mu \nu
}^{SU_{c}\left( 3\right) },\mathcal{F}_{\mu \nu }^{U_{em}\left(
1\right) })$, one is left with a physics quite similar to the one of
the AQHE of the delocalized electrons of \emph{2D} graphene.
Therefore, one expects the light quark's dynamics to show as well an
anomalous quantum Hall type phenomenon\textrm{\footnote{%
By QHE, we mean a quantized conductivity of the system following
from the existence of a discrete energy spectrum and a discrete
filling factor due to the background field.}} in the presence of a
constant $\mathcal{F}_{\mu \nu
}^{G}$. This tensor reads generally in terms of the gauge potential as $%
\partial _{\mu }A_{\nu }^{G}-\partial _{\nu }A_{\mu }^{G}+\left[ A_{\mu
}^{G},A_{\nu }^{G}\right] $; but for explicit computations, we will
mainly focus on the abelian part of the gauge symmetry G. \newline
Because of the uniformity condition$\ \frac{\partial }{\partial x}\mathcal{F}%
_{\mu \nu }^{G}=0$, and also by restricting $\mathcal{F}_{\mu \nu
}^{G}$ to take values in the Cartan subalgebra $U_{c}^{2}\left(
1\right) \times
U_{em}\left( 1\right) \subset G$, the abelian part of the gauge potential $%
A_{\mu }^{I}\left( x\right) $ that obeys the Lorentz condition
$\partial
_{\mu }A^{\mu I}=0$ reads, up to irrelevant numbers, as follows $\frac{1}{2}%
\mathcal{F}_{\mu \nu }^{I}x^{\nu }$. Notice by the way that unlike
\emph{2D} graphene, the underlying space time of \emph{4D} lattice
\emph{QCD} has four real euclidian dimensions leading to several
possibilities for the allowed directions of $\mathcal{F}_{\mu \nu
}$. This richness, which has also a physical interpretation in terms
of interactions (see section 4), may be fixed by looking for
configurations that permit the diagonalization of the \emph{BC}-
hamiltonian $H_{BC}$ which acts on $SO\left( 4\right) $ spinorial
states $\left\vert \Psi _{E}\right\rangle $ like $H_{BC}\left\vert
\Psi
_{E}\right\rangle =E\left\vert \Psi _{E}\right\rangle $ with $H_{BC}=\frac{1%
}{i}\gamma ^{\mu }\left( \partial _{\mu }-i\frac{Q}{c}A_{\mu
}\right) $ and where the four matrices $\gamma ^{\mu }$ are the
usual Dirac 4$\times $4 matrices in the 4D euclidian space. By
choosing the constant background fields like,
\begin{equation}
\begin{tabular}{llll}
&  &  &  \\
$\mathcal{F}_{\mu \nu }^{em}$ & $=$ & $\mathcal{B}\varepsilon _{\mu \nu 34}+%
\mathcal{E}\varepsilon _{12\mu \nu }\quad ,\quad \det
\mathcal{F}_{\mu \nu
}^{em}=\mathcal{B}^{2}\times \mathcal{E}^{2}$ &  \\
&  &  &  \\
$\mathcal{F}_{\mu \nu }^{QCD}$ & $=$ & $\dsum\limits_{I=1}^{2}h_{I}\mathcal{F%
}_{\mu \nu }^{I}+\dsum\limits_{su_{3}\text{ roots }\alpha }E^{-\alpha }%
\mathcal{F}_{\mu \nu }^{\alpha }=0$ & ,%
\end{tabular}
\label{FF}
\end{equation}%
breaking down the $SO\left( 4\right) $ symmetry of the euclidian $\mathbb{R}%
^{4}$ down to $SO\left( 2\right) \times SO\left( 2\right) $, one
ends with a diagonal form of the squared operator $H_{BC}^{2}$; as
well as remarkable factorized relations that allow to perform the
explicit computation of the
exact hamiltonian spectrum. Notice that, viewed from the 4D space time with $%
SO\left( 4\right) $ symmetry, the constants $\mathcal{B}$ and
$\mathcal{E}$, appearing in the above relations, are "magnetic" and
"electric" components of $\mathcal{F}_{\mu \nu }^{U_{em}\left(
1\right) }$ respectively normal to the x$^{1}$-x$^{2}$ and
x$^{3}$-x$^{4}$ real planes of $\mathbb{R}^{4}$.
However, from a (1+4)-dimensional space time\textrm{\footnote{%
In 1+4 dimensions, the antisymmetric tensor $\mathcal{F}_{MN}$ has
\emph{10} components; \emph{6} of them given by $\mathcal{F}_{\mu
\nu }$ are magnetic type and the other \emph{4} ones given by
$\mathcal{F}_{5\mu }$ are of
electric type.}} with $SO\left( 1,4\right) $ isotropy symmetry, both of $%
\mathcal{B}$ and $\mathcal{E}$ behave as magnetic components that
couple to left $\Psi _{L}$ and right $\Psi _{R}$ handed fermions
respectively; and so lead to QHE phenomenon in four dimensions. This
\emph{5D} interpretation will be discussed with some details in
section 3 and in conclusion. With the above diagonal choice of
$\mathcal{F}_{\mu \nu }^{_{em}}$, which contains as particular cases
the singular limit $\det \mathcal{F}_{\mu \nu }^{_{em}}=0$
describing chiral configurations with solely $\Psi _{L}$ or $\Psi
_{R}$, we find, amongst others and besides the \emph{5D}
interpretation of the BC model, the two following features:

\ \ \ \ \newline (\textbf{1}) the energy spectrum the BC fermions in
the constant background fields (\ref{FF}) is discrete provided the
$\mathcal{F}_{\mu \nu }^{em}$ tensor is non degenerate; that is as
far as the product $\mathcal{B}\times \mathcal{E}\neq 0$. In this
case, the energy spectrum is given by
\begin{equation}
E_{n,m}^{\pm }\left( \varpi ,\varpi ^{\prime }\right) =\pm \hbar \sqrt{%
n\varpi ^{2}+m\varpi ^{\prime 2}}\text{, \qquad }n,m\geq 0,
\label{BE}
\end{equation}%
with the oscillator frequencies $\varpi ^{2}=\frac{2\left\vert Q\mathcal{B}%
\right\vert }{c}$, $\varpi ^{\prime 2}=\frac{2\left\vert Q\mathcal{E}%
\right\vert }{c}$; and where $Q=qe$ stand for the electric
fractional charge of the quark $u_{\alpha }\left( \mathbf{x}\right)
$ or the quark $d_{\alpha }\left( \mathbf{x}\right) $ living at each
of the two valleys of the BC model. Clearly the appearance of the
frequency $\varpi ^{\prime }$ is due to euclidian nature of the 4D
lattice and to the underlying \emph{5D} interpretation where
$\mathcal{E}$ is thought of as an external "magnetic" field that
couple to right handed $\Psi _{R}$. In the singular limit $\det
\mathcal{F}_{\mu \nu }^{_{em}}=0$, for instance in case where $\mathcal{E}%
\rightarrow 0$, the above energy spectrum gets modified as
\begin{equation*}
E_{n}^{\pm }\left( \varpi ,k_{z}\right) =\pm \hbar \varpi \sqrt{n+\frac{%
k_{z}^{2}+k_{\tau }^{2}}{\varpi ^{2}}},
\end{equation*}%
with $\hbar k_{i}$ being the momenta of $\Psi _{R}$ along the
\textit{i-th} direction. These energies define a family of pairs of
opposite paraboloids separated by the gaps $\Delta E_{n}=$
$2\sqrt{n\varpi ^{2}+k_{z}^{2}+k_{\tau }^{2}}$, and touch on the
fundamental state $n=0,$ $k_{z}=0,$ $k_{\tau }=0$ leading to a zero
gap system; see also fig \ref{QH} for illustration.

\ \ \ \newline (\textbf{2}) The \emph{BC} fermion's filling factor
$\nu _{BC}$ of the anomalous quantum Hall effect induced by the
background fields (\ref{FF}) in the band energy $0\leq
E_{n,m}^{+}\leq E_{N,M}^{+}$, with $N$, $M$ positive
integers, is given by,%
\begin{equation}
\nu _{BC}=k_{V}\times N_{c}\times g_{L}\times g_{R}\times
\frac{\left( 2N+1\right) \left( 2M+1\right) }{2}  \label{FR}
\end{equation}%
where $k_{V}$ is the number of Dirac valleys, {\normalsize which
takes the value }$k_{V}=2${\normalsize \ for minimally doubled
fermions including BC and KW ones; }$k_{V}=16${\normalsize \ for
naive fermions and }$k_{V}=4$ {\normalsize for staggered fermions};
$N_{c}$ the quark's color number which is equal to 3; and
$g_{L}=g_{R}=2$ the number of spin polarizations of the left and
right handed fermions of the $SO\left( 4\right) $ spinor. Notice
that in the case where, in addition to $\mathcal{F}_{\mu \nu }^{_{em}}=%
\mathcal{B}\varepsilon _{\mu \nu 34}+\mathcal{E}\varepsilon _{12\mu
\nu }$, we also have $\mathcal{F}_{\mu \nu }^{SU_{c}\left( 3\right)
}\neq 0$ along the two directions of the Cartan subalgebra of the
SU$_{C}\left( 3\right) $ gauge symmetry, BC fermions coupled to the
background fields are described by \emph{6} oscillators (\emph{2}
for each color; i.e: for each $\Psi _{L}^{c}$ and $\Psi _{R}^{c}$,
with $c=1,2,3$) and the above filling factor generalizes as follows,
\begin{equation}
\nu _{BC}^{gen}=k_{V}\times g_{L}\times g_{R}\times \frac{1}{2}%
\dprod\limits_{i=1}^{6}\left( 2N_{i}+1\right) \text{.}  \label{FL}
\end{equation}%
The presentation is as follows. In section 2, we review briefly the
anomalous quantum Hall effect (AQHE) of graphene; this section aims
also to
revisit useful aspects of graphene using relativistic field theory in $%
\left( 1+2\right) $\emph{D }; and also to describe our approach on a
simple system. In section 3, we study the minimally doublet fermions
in a constant background field by focusing on BC fermions. In
particular we study the BC model of 4D\ lattice QCD; first as a
lattice field theory on the hyperdiamond; and second as a $\left(
1+4\right) $\emph{D}\ extension of the graphene near the Dirac
points. In section 4, we study the algebra of the gauge covariant
derivatives and its highest weight representations. These
representations are used in section 5 to determine the spectrum of
the BC-hamiltonian in presence of background fields; and determine
the filling factor of the AQHE of the BC fermions. We also give the
relation with the spectral flow hamiltonian considered in
\textrm{\cite{4K,4L,2BB,2BC,4M}}. Last section is devoted to
conclusion and comments.\ In the appendix A and appendix B, we
develop further the link between our study and the index theorem.

\section{AQHE in 3D relativistic systems: case of \emph{2D} graphene}

In this section, we study the anomalous quantum Hall effect of a
relativistic fermionic system described by the $\left( 1+2\right) $
Dirac equation in a constant magnetic background field. This
concerns a particular QED$_{3}$ model where the electromagnetic
field strength $F_{\mu \nu }$ takes a constant value with direction
normal to the 2D space surface where live the sheet of graphene. We
assume that the magnitude of the external field B is bounded as
follows \emph{14} \emph{Tesla} $\leq $ $\left\vert B\right\vert \leq
$ \emph{20} \emph{Tesla} so that the Zeeman coupling can be
ignored\textrm{\ \cite{5F}}.

\subsection{Dirac equation in 3D}

First consider the $\left( 1+2\right) $ space time
$\mathbb{R}^{1,2}$ with
local coordinates $x^{\mu }=\left( t,x,y\right) $, embedded in the usual $%
\left( 1+3\right) $ dimensional space time $\mathbb{R}^{1,3}$
parameterized by $X^{M}=\left( x^{\mu },z\right) $; then focus on
the Dirac equation of a fermionic particle, described by the complex
field doublet $\psi ^{a}=\left( \phi ,\chi \right) $ living in
$\mathbb{R}^{1,2}$, in the presence of an
external constant magnetic field $B$ taken along the z-direction of $\mathbb{%
R}^{1,3}$,
\begin{equation}
\sum_{\mu =0}^{2}\sum_{b=1}^{2}i\sigma _{ab}^{\mu }\left( D_{\mu
}\psi ^{b}\right) =0.
\end{equation}%
In this system of equations, the hermitian $2\times 2$ matrix $\mathcal{D}%
=i\sigma _{ab}^{\mu }D_{\mu }$ with $iD_{\mu }=i\left( \partial _{\mu }-i%
\frac{e}{c}A_{\mu }\right) $ and $\left( iD_{\mu }\right)
^{+}=iD_{\mu }$,
is the gauged Dirac operator in $\left( 1+2\right) $ spacetime dimensions, $%
\sigma ^{0}=I_{2}$ the unity matrix and $\sigma ^{i}=\left( \sigma
^{1},\sigma ^{2}\right) $ the usual $2\times 2$ Pauli matrices
obeying the 2D Clifford algebra $\sigma ^{i}\sigma ^{j}+\sigma
^{j}\sigma ^{i}=2\delta ^{ij}$ as well as the $SU\left( 2\right) $
commutation relations type $\left[ \sigma ^{1},\sigma ^{2}\right]
=2i\sigma ^{3}$ giving the $\pm $ chiralities of the component
fields $\phi $ and $\chi $. Moreover, $A_{\mu }$ is the
gauge potential which in the present case reads as $\frac{1}{2}\mathcal{F}%
_{\mu \nu }x^{\nu }$ where the constant field strength tensor
\begin{equation}
\mathcal{F}_{\mu \nu }=\frac{c}{ie}\left[ D_{\mu },D_{\nu }\right]
\end{equation}%
\ is given by the magnetic field $\mathcal{B}\varepsilon _{0\mu \nu
3}$ along the z-direction. Here, the rank 4 tensor $\varepsilon
_{\alpha \mu \nu \beta }$ is the completely antisymmetric $SO\left(
1,3\right) $ invariant tensor; and the restricted $\varepsilon
_{0\mu \nu 3}\equiv \varepsilon _{\mu \nu }$ is the antisymmetric
$SO\left( 2\right) $ invariant $2\times 2$ matrix. Substituting
$\mathcal{F}_{\mu \nu }=\mathcal{B}\varepsilon _{\mu \nu }$, we see
that the linear gauge potential lives in the x-y plane with
components as given below,%
\begin{equation}
\begin{tabular}{ll}
$A_{1}=+\frac{\mathcal{B}}{2}y,$ & $A_{0}=0$, \\
$A_{2}=-\frac{\mathcal{B}}{2}x,$ & $\partial _{\mu }A^{\mu }=0.$%
\end{tabular}%
\end{equation}%
From the above relations, we learn also that $\left[ D_{1},D_{2}\right] =i%
\frac{e}{c}\mathcal{B}$ showing amongst others that $\mathcal{B}$
acts as a deformation parameter on the free spectrum of the Dirac
operator; and the algebra of the gauge covariant derivatives, is up
to a scale factor, the well known Heisenberg algebra of the harmonic
oscillator; i.e $\left[ A,A^{\dagger }\right] =\hbar I$.
Furthermore, factorizing the time
dependence of the wave function by summing over all possible frequencies as%
\begin{equation}
\psi ^{a}\left( t,x,y\right) =\int_{-\infty }^{+\infty
}\frac{dE}{2\pi \hbar }e^{-\frac{i}{\hbar }Et}\psi _{E}^{a}\left(
x,y\right)
\end{equation}%
or equivalently by using discrete spectrum notation like%
\begin{equation}
\psi ^{a}\left( t,x,y\right) =\dsum\limits_{n=-\infty }^{+\infty }e^{-\frac{i%
}{\hbar }E_{n}t}\psi _{n}^{a}\left( x,y\right) ,  \label{WF}
\end{equation}%
with $E_{n}=\hbar \omega _{n}$; and $\psi _{+\left\vert n\right\vert
}^{a}\left( x,y\right) $, $\psi _{-\left\vert n\right\vert
}^{a}\left( x,y\right) $ the wave functions associated with positive
and negative frequencies; then using the fact that scalar potential
$A_{0}=0$, it is convenient to rewrite the Dirac equation of $\psi
_{n}^{a}$ as follows
\begin{equation}
\left(
\begin{array}{cc}
0 & i\left( D_{1}-iD_{2}\right) \\
i\left( D_{1}+iD_{2}\right) & 0%
\end{array}%
\right) \left(
\begin{array}{c}
\phi _{n} \\
\chi _{n}%
\end{array}%
\right) =\omega _{n}\left(
\begin{array}{c}
\phi _{n} \\
\chi _{n}%
\end{array}%
\right) .  \label{ME}
\end{equation}%
The solutions of this matrix equation give the possible frequencies
$\omega _{n}$ and the corresponding wave function doublets $\left(
\phi _{n},\chi _{n}\right) $, with the integer $n$ refereing to the
quantized values of energy and momentum due to the background field
$\mathcal{B}$. To that end, acting on the above 2$\times $2 matrix
equation (\ref{ME}) by the Dirac operator $i\sum \sigma ^{l}D_{l}$,
we can bring it into the following
diagonal one%
\begin{equation}
\left(
\begin{array}{cc}
D_{-}D_{+} & 0 \\
0 & D_{+}D_{-}%
\end{array}%
\right) \left(
\begin{array}{c}
\phi _{n} \\
\chi _{n}%
\end{array}%
\right) =\omega _{n}^{2}\left(
\begin{array}{c}
\phi _{n} \\
\chi _{n}%
\end{array}%
\right)  \label{DD}
\end{equation}%
where we have set $D_{\pm }=i\left( D_{1}\pm iD_{2}\right) $ with
the property $\left( D_{\pm }\right) ^{+}=D_{\mp }$. We also have
\begin{equation}
\begin{tabular}{lllll}
$D_{+}=i\left( 2\frac{\partial }{\partial \bar{u}}-\frac{e\mathcal{B}}{2c}%
u\right) $ & , & $u=x+iy,$ & $\frac{\partial }{\partial u}=\frac{1}{2}\frac{%
\partial }{\partial x}-\frac{i}{2}\frac{\partial }{\partial y}$ & , \\
$D_{-}=i\left( 2\frac{\partial }{\partial u}+\frac{e\mathcal{B}}{2c}\bar{u}%
\right) $ & , & $\bar{u}=x-iy,$ & $\frac{\partial }{\partial \bar{u}}=\frac{1%
}{2}\frac{\partial }{\partial x}+\frac{i}{2}\frac{\partial }{\partial y}$ & ,%
\end{tabular}%
\end{equation}%
together with the angular momentum $L_{z}$ around the z-axis reading like%
\begin{equation}
L_{z}=\left( u\frac{\partial }{\partial u}-\bar{u}\frac{\partial
}{\partial
\bar{u}}\right) =\frac{1}{i}\left( x\frac{\partial }{\partial y}-y\frac{%
\partial }{\partial x}\right) ,
\end{equation}%
and acting on $D_{\pm }$ and $D_{-}D_{+}$ as follows
\begin{equation}
\begin{tabular}{llll}
$\left[ L_{z},D_{\pm }\right] =\pm D_{\pm }$ & $,$ & $\left[ L_{z},D_{-}D_{+}%
\right] =0$ & .%
\end{tabular}%
\end{equation}%
The relation $\left[ L_{z},D_{-}D_{+}\right] =0$ shows that the wave
functions carry, in addition to energy, also an angular momentum
charge that we skip below for simplicity. Notice moreover that the
$D_{\pm }$ operators satisfy the commutation relation $\left[
D_{-},D_{+}\right] =\frac{2e}{c}B$ which, up on using the
normalization
\begin{equation}
D_{-}=A\sqrt{\frac{2e}{c}\mathcal{B}},\qquad D_{+}=A^{\dagger }\sqrt{\frac{2e%
}{c}\mathcal{B},}
\end{equation}%
with positive $B$, can be put in the usual Heisenberg form
\begin{equation}
\begin{tabular}{llll}
$\left[ A,A^{\dagger }\right] =I$ & , & $\left[ A,A\right] =\left[
A^{\dagger },A^{\dagger }\right] =0$ & ,%
\end{tabular}%
\end{equation}%
describing the algebra of the quantum harmonic oscillator with
$A^{\dagger }$
the creation operator, $A$ the annihilation operator; and $\hbar I$ (with $%
\hbar $ set to one) the diagonal one that captures the energy
quantum number $n$. The highest weight representations of this
algebra are well known; they are given by semi infinite dimensional
vector spaces generated by the normalized basis vectors $\left\{ \xi
_{\pm n}\left( x,y\right) =|\xi _{\pm n}>\right\} _{n\geq 0}$ based
on the highest weight vector $|\xi _{0}>$
obeying the conditions%
\begin{equation}
\begin{tabular}{llll}
$A|\xi _{0}>=0$ & , & $\hbar I|\xi _{0}>=\hbar |\xi _{0}>$ & ,%
\end{tabular}%
\end{equation}%
solved generally like $\xi _{0}\left( u,\bar{u}\right) \sim u^{l}e^{-\frac{e%
\mathcal{B}}{4c}u\bar{u}}$ with $l$ the value of the angular
momentum which we set below to zero. The other states of the
representation are given by successive application of the creation
operator $A^{\dagger }$ as given below,
\begin{equation}
\begin{tabular}{llll}
$A^{\dagger }|\xi _{n}>$ & $=\sqrt{n+1}|\xi _{n+1}>$ & , & $n\geq 0$ \\
$A|\xi _{n}>$ & $=\sqrt{n}|\xi _{n-1}>$ & , & $\mathcal{N}=A^{\dagger }A$%
\end{tabular}%
\end{equation}%
with $|\xi _{-1}>=0$ as well as
\begin{equation}
\begin{tabular}{ll}
$\mathcal{N}|\xi _{n}>=n|\xi _{n}>,$ & $<\xi _{n}|\xi _{m}>=\delta _{nm}.$%
\end{tabular}%
\end{equation}%
Substituting the operators $D_{\pm }$ in terms of their expression
using $A$ and $A^{\dagger }$, we can split the matrix equation
(\ref{DD}) in two
eigenstate equations as follows%
\begin{equation}
\begin{tabular}{llll}
$\left( A^{+}A+I\right) $ $\phi _{n}$ & $=$ & $\frac{\omega _{n}^{2}c}{2e%
\mathcal{B}}$ $\phi _{n}$ & , \\
$A^{+}A$ $\chi $ & $=$ & $\frac{\omega _{n}^{2}c}{2e\mathcal{B}}$
$\chi _{n}$
& .%
\end{tabular}%
\end{equation}%
The wave functions of these relations are naturally solved by the
basis states $\left\{ |\xi _{n}>\right\} $ by taking $\phi _{n}$ and
$\chi _{n}$
like%
\begin{equation}
\left(
\begin{array}{c}
\phi _{n} \\
\chi _{n}%
\end{array}%
\right) =\left(
\begin{array}{c}
|\xi _{n-1}> \\
|\xi _{n}>%
\end{array}%
\right) ,
\end{equation}%
describing two neighboring polarized particles. The energies follow
from the
relation between the squared frequencies $\omega _{n}^{2}=\frac{E_{n}^{2}}{%
\hbar ^{2}}$ and the integer $n$. We have%
\begin{equation}
\begin{tabular}{llll}
$\omega _{n}^{2}=\frac{2eB}{c}n$ & , & $E_{n}^{\pm }=\pm \sqrt{\frac{2\hbar e%
\mathcal{B}}{c}\left\vert n\right\vert }$ & ,%
\end{tabular}
\label{EN}
\end{equation}%
with $E_{n}^{+}$ and $E_{n}^{-}$ respectively associated with the waves $%
\psi _{+\left\vert n\right\vert }^{a}\left( x,y\right) $ and $\psi
_{-\left\vert n\right\vert }^{a}\left( x,y\right) $ of eq(\ref{WF}).
Notice
that for $n=0$, we have a zero mode with negative chirality%
\begin{equation}
\left(
\begin{array}{c}
0 \\
\chi _{0}%
\end{array}%
\right)  \label{CZ}
\end{equation}

\subsection{Filling factor}

As was discovered in 1980, the Hall conductivity $\sigma _{xy}$ of a \emph{2D%
} electron gas in a strong transverse magnetic field develops
plateaus at
values quantized in units of $\frac{e^{2}}{h}$; i.e $\sigma _{xy}=\nu \frac{%
e^{2}}{h}$ with $\nu =\frac{N_{f}}{N_{\phi }}$ standing for the
filling factor \textrm{\cite{4IB,1R}}. In the case of 2D graphene,
the filling factor $\nu _{graph}$ of the QHE is given by the ratio
of the number $N_{F}$ of particle states with respect the number of
quantum fluxes $N_{\phi }=\int_{S^{2}}\mathcal{F}_{2}$ where the
2-form $\mathcal{F}_{2}$ is given by $\mathcal{B}dx\wedge dy$. We
have,
\begin{equation}
\nu _{graph}=\frac{N_{f}}{N_{\phi }}.
\end{equation}%
A direct way to get this factor is to compute the number $N_{F}$ per
a unit quantum flux $N_{\phi }=1$. For that purpose we proceed as
follows: First recall that the $\psi _{+\left\vert n\right\vert }$
and $\psi _{-\left\vert
n\right\vert }$ components of the expansion of the quantum wave function $%
\psi =\sum_{n}\psi _{n}$ with $\psi _{n}\left( t,x,y\right)
=e^{-i\omega _{n}t}\psi _{n}\left( x,y\right) $ as in eq(\ref{WF})
describe respectively
particles and holes that live on the energy levels $E_{n}^{+}$ and $%
E_{n}^{-} $ of eq(\ref{EN}). Second, the total number of $N_{F}$
polarized
particles and $N_{F}$ holes, per unit quantum flux, that fill the band energy%
\begin{equation}
0\leq E_{n}^{2}\leq E_{N}^{2},\qquad E_{N}^{-}\leq E_{n}\leq
E_{N}^{+},
\end{equation}%
with $E_{N}^{\pm }=\pm \sqrt{\frac{2\hbar e\mathcal{B}}{c}\left\vert
N\right\vert }$ is, because of symmetry, precisely $2N_{F}$; and is
given by the relation
\begin{equation}
2N_{F}=\int_{E_{N}^{-}\leq E_{n}\leq E_{N}^{+}}dtd^{2}x\left\vert
\psi \left( t,x,y\right) \right\vert ^{2}\   \label{N}
\end{equation}%
together with the expansion%
\begin{equation}
\psi \left( t,x,y\right) =\dsum\limits_{-N}^{+N}e^{-i\omega
_{n}t}\psi _{n}\left( x,y\right) .  \label{E}
\end{equation}%
Putting back (\ref{E}) into (\ref{N}) and using the orthogonality property $%
\int d^{2}x$ $\bar{\psi}_{n}\left( x,y\right) \psi _{m}\left( x,y\right) $ $%
=\delta _{n,m},$ the relation (\ref{N}) reduces to
\begin{equation}
2N_{F}=\dsum\limits_{-N}^{+N}1=\left( 2N+1\right) .
\end{equation}%
From the above relations, we learn that the positive energy values $%
E_{n}^{+} $ are associated with the wave functions indexed by
positive
integers $n\geq 0$ namely $\psi _{+n}^{a}\left( t,x,y\right) =$ $e^{-\frac{i%
}{\hbar }E_{n}^{+}t}\psi _{+n}^{a}\left( x,y\right) $; they
correspond, in the language of condensed matter physics, to the
conducting band where live particle states. The negative energy
values $E_{n}^{-}$ are associated with the wave functions indexed by
negative integers $n=-\left\vert n\right\vert
\leq 0$, $\psi _{-\left\vert n\right\vert }^{a}\left( t,x,y\right) =e^{-%
\frac{i}{\hbar }E_{n}^{-}t}\psi _{-\left\vert n\right\vert
}^{a}\left( x,y\right) $; they are associated with the valence band
where live the holes. In the particular case $n=0$, both energies
$E_{n}^{\pm }$ vanish and the conducting and valence bands touch. So
the fundamental state $\psi
_{0}^{a}$ with zero energy would be filled by $\frac{1}{2}$ particle and $%
\frac{1}{2}$ hole; this property can be viewed as another statement
of the chiral anomaly.\newline Therefore the number of polarized
particles filling the band energy $0\leq E_{n}^{+}\leq
\sqrt{\frac{2\hbar eB}{c}\left\vert N\right\vert }$ is equal to
$N+\frac{1}{2}$. So, by taking into account the fact that electrons
has two spin polarizations, the general expression of the filling
factor $\nu
_{graph}$ reads as follow:%
\begin{equation}
\nu _{graph}=2k_{V}\left( N+\frac{1}{2}\right) \equiv 4N+2,
\end{equation}%
where $k_{V}$ stands for the number of Dirac valleys which is equal
to $2$ in the case of graphene.

\section{Bori\c{c}i-Creutz fermions in background fields}

In this section, we extend known results on \emph{2D} graphene to BC
fermions of \emph{4D} lattice QCD; but the analysis applies
straightforwardly\ to KW and naive fermions. First, we recall the
\emph{BC} fermions and its two Dirac valleys as being a simple model
for lattice QCD simulations. Then we study some remarkable
properties of the Dirac equation in the background field (\ref{FF}).
The explicit computation of the spectrum of \emph{BC} fermions in
background fields will be given in next sections.

\subsection{BC model and 4D graphene}

Bori\c{c}i-Creutz fermions \textrm{\cite{3A}} is a simple four
dimensional lattice QCD model for studying and simulating the
interacting dynamics of
the two light quarks up $u_{\alpha }\left( \mathbf{x}\right) $ and down $%
d_{\alpha }\left( \mathbf{x}\right) $; see ref.\textrm{\cite{2B}}
for other specific features of BC fermions; in particular the
explicit derivation of the $SU\left( 2\right) $
flavor\footnote{\textrm{To be precise the species in lattice
fermions should be treated as point splitting fields as proposed in
\cite{2BB} and used extensively in \cite{2BC}}.} symmetry $\left(
u_{\alpha },d_{\alpha }\right) $. The BC model\textrm{\ }can be
derived as a particular model resulting from the following general
euclidian lattice action
\begin{equation}
\mathcal{S}_{BC}=\frac{1}{\mathrm{a}}\sum_{\mathbf{x}_{i}\mathbf{\in }%
\mathbb{L}}\left( \sum_{l=1}^{5}\Psi _{\mathbf{x}_{i}}^{+}\gamma
^{\mu
}\Omega _{\mu }^{l}\Psi _{\mathbf{x}_{i}\mathbf{+v}_{l}}-\Psi _{\mathbf{x}%
_{i}}^{+}\gamma ^{\mu }\bar{\Omega}_{\mu }^{l}\Psi _{\mathbf{x}_{i}\mathbf{-v%
}_{l}}\right) ,
\end{equation}%
where $\mathbb{L}$ is the \emph{4D} hyperdiamond with lattice
parameter \textrm{a}; $\Psi _{\mathbf{x}_{i}}$ is a 4D euclidian
Dirac spinor at the site $x_{i}$; $\gamma ^{\mu }$ the usual
$4\times 4$ Dirac matrices to be specified later; $\mathbf{v}_{l}$
are five relative vectors parameterizing the first nearest
neighbors; and where $\Omega _{\mu }^{l}$ is a $4\times 5$ tensor
that reads like \textrm{\cite{5A}},
\begin{equation}
\Omega _{\mu }^{l}=\left(
\begin{array}{ccccc}
-\frac{1}{2} & -\frac{1+2i}{4} & \frac{1}{4} & \frac{1}{4} & \frac{1}{4} \\
-\frac{1}{2} & \frac{1}{4} & -\frac{1+2i}{4} & \frac{1}{4} & \frac{1}{4} \\
-\frac{1}{2} & \frac{1}{4} & \frac{1}{4} & -\frac{1+2i}{4} & \frac{1}{4} \\
-\frac{1}{2} & \frac{1}{4} & \frac{1}{4} & \frac{1}{4} & -\frac{1+2i}{4}%
\end{array}%
\right) .
\end{equation}%
Expanding the Dirac spinors in Fourier modes, we get
$\mathcal{S}_{BC}=\int
\frac{d^{4}\mathbf{k}}{\left( 2\pi \right) ^{4}}\Psi _{\mathbf{k}}^{+}i%
\mathcal{D}_{BC}\Psi _{\mathbf{k}}$ where the $4\times 4$ matrix $\mathcal{D}%
_{BC}$ is the \emph{BC} operator in the reciprocal space; it reads
as
follows,%
\begin{equation}
\begin{tabular}{ll}
$\mathcal{D}_{BC}\text{ }\sim \text{ }\frac{1}{a}\dsum\limits_{\mu =1}^{4}%
\mathrm{\gamma }^{\mu }\sin ak_{\mu }-\frac{1}{a}\dsum\limits_{\mu =1}^{4}%
\mathrm{\gamma }^{\mu }\cos ak_{\mu }+\frac{1}{a}\dsum\limits_{\mu
=1}^{4}\Gamma \cos ak_{\mu }-\frac{2}{a}\Gamma $ & ,%
\end{tabular}
\label{DI}
\end{equation}%
with
\begin{equation}
\Gamma =\frac{1}{2}\left( \gamma ^{1}+\gamma ^{2}+\gamma ^{3}+\gamma
^{4}\right) .
\end{equation}%
From these expressions, one can check that the operator
$\mathcal{D}_{BC}$
has two zero modes given by the two following four component wave vectors $%
k_{\mu }$,%
\begin{equation}
\begin{tabular}{lllll}
$\left( i\right) $ & : $\ \ K$ & $=\left(
k_{1},k_{2},k_{3},k_{4}\right) $ &
$=\left( 0,0,0,0\right) $ & , \\
$\left( ii\right) $ & : $\ \ K^{\prime }$ & $=\left(
k_{1},k_{2},k_{3},k_{4}\right) $ & $=\left( \frac{\pi }{2a},\frac{\pi }{2a},%
\frac{\pi }{2a},\frac{\pi }{2a}\right) $ & .%
\end{tabular}
\label{ZM}
\end{equation}%
The first zero mode is associated with the quark $\tilde{u}_{\alpha
}\left( \mathbf{k}\right) $; and the second with the quark
$\tilde{d}_{\alpha
}\left( \mathbf{k}\right) ,$ these fields are just the Fourier transform of $%
u_{\alpha }\left( \mathbf{x}\right) $ and $d_{\alpha }\left( \mathbf{x}%
\right) $ respectively. Expanding eq(\ref{DI}) near the zero modes (\ref{ZM}%
-i) and (\ref{ZM}-ii), one gets at the first order in $k_{\mu }$,
\begin{equation}
\mathcal{D}_{BC}\text{ }\simeq \dsum\limits_{\mu =1}^{4}\mathrm{\gamma }%
^{\mu }k_{\mu }\rightsquigarrow \dsum\limits_{\mu =1}^{4}\mathrm{\gamma }%
^{\mu }\frac{\partial }{i\partial x^{\mu }},
\end{equation}%
which is precisely the Dirac operator of a free fermion. As such,
near each one of the two Dirac points the \emph{BC} model reads in
the continuous
space as follows%
\begin{equation}
\mathcal{S}_{BC}=\int_{\mathbb{R}^{4}}d^{4}\mathbf{x}\Psi ^{+}\left( \mathbf{%
x}\right) \mathrm{\gamma }^{\mu }\frac{\partial \Psi \left( \mathbf{x}%
\right) }{\partial x^{\mu }},  \label{SBC}
\end{equation}%
with $\Psi \left( \mathbf{x}\right) $ standing for the quarks
$u\left( \mathbf{x}\right) $ and $d\left( \mathbf{x}\right) $
respectively given by integrals over momentum hyper-balls
$\mathbb{B}$ centered at $\mathbf{K}$ and $\mathbf{K}^{\prime }$ as
reported below; see \textrm{\cite{2B} }for technical details,
\begin{equation}
\begin{tabular}{ll}
$u\left( \mathbf{x}\right) =\dsum\limits_{\mathbf{y}}\int_{\mathbb{B}_{%
\mathbf{K}}}\frac{d^{4}\mathbf{k}}{\left( 2\pi \right) ^{4}}e^{i\mathbf{k}%
\left( \mathbf{x}-\mathbf{y}\right) }u\left( \mathbf{y}\right) $ & , \\
$d\left( \mathbf{x}\right) =\dsum\limits_{\mathbf{y}}\int_{\mathbb{B}_{%
\mathbf{K}^{\prime }}}\frac{d^{4}\mathbf{k}}{\left( 2\pi \right) ^{4}}e^{i%
\mathbf{k}\left( \mathbf{x}-\mathbf{y}\right) }d\left(
\mathbf{y}\right) $ &
.%
\end{tabular}%
\end{equation}%
Notice that the way \textrm{this the above relation is a naive
definition of the flavors; a more concise way to do is that done in
\cite{2BB}. Although not used here below, this method is based on
point splitting fields in the reciprocal space. }For later use, it
is interesting to notice as well the following remarkable
properties: \newline First, one has to think about the field $\Psi
\left( x\right) $ as the zero
mode of the hamiltonian equation%
\begin{equation}
\left\langle x\right\vert H_{BC}\left\vert \Psi _{E}\right\rangle
=E\Psi _{E}\left( \mathbf{x}\right) \;\;\;,\;\;\;\left\langle
x\right\vert H_{BC}\left\vert \Psi _{0}\right\rangle =0\;\;\;,
\label{HBC}
\end{equation}%
with hamiltonian\textrm{\footnote{%
Notice that $H_{BC}$ has been defined as an invariant under the
$SO\left( 4\right) $ symmetry that follows from the continuum limit
of the hyperdiamond. If one insists on defining the hamiltonian in
terms the translation generator along the 4-th direction, one has to
break $SO\left( 4\right) $ down to $SO\left( 3\right) $; and ends
amongst others either with $\det \mathcal{F}_{\mu \nu }=0$ or with a
quantization condition on the background fields $\mathcal{B}$ and
$\mathcal{E}$.}}
\begin{equation}
H_{BC}=\mathrm{\gamma }^{\mu }\frac{\partial }{i\partial x^{\mu }}.
\end{equation}%
To interpret eq(\ref{HBC}) as an equation of motion, one has to
modify the
action $\mathcal{S}_{BC}$ (\ref{HBC}) by adding an extra term of the form $%
\frac{E}{i}$ $\Psi _{E}^{+}\left( x\right) \Psi _{E}\left( x\right)
$.
\newline
Second, since the parameter $E$ can take any real value, we will
also think about the Dirac spinor $\Psi _{E}\left( x\right) $ as
describing modes of a
spinor field living on a \emph{5D} space time as given below%
\begin{equation}
\mathbf{\Psi }\left( \mathbf{x},t\right) =\int \frac{dE}{2\pi \hbar }e^{%
\frac{i}{\hbar }Et}\Psi _{E}\left( \mathbf{x}\right) ,  \label{311}
\end{equation}%
so that the \emph{4D} action (\ref{HBC}) looks as resulting from the
following five dimensional one
\begin{equation}
\mathcal{S}_{5D}=\int_{\mathbb{R}^{1+4}}dtd^{4}\mathbf{x}\left[
\mathbf{\Psi
}^{+}\left( \mathbf{x,}t\right) \left( \dsum\limits_{\mu =1}^{4}\mathrm{%
\gamma }^{\mu }\frac{\partial }{\partial x^{\mu }}-\frac{\partial
}{\partial t}\right) \mathbf{\Psi }\left( \mathbf{x,}t\right)
\right] ,
\end{equation}%
where $t$ stands for the five dimensional coordinate; and which
could be interpreted as a "time" coordinate encoding non zero energy
deformations near the Dirac points. This \emph{5D} interpretation
will be developed with further details later on;\ it will be also
linked with the so called spectral flow hamiltonian given in
sub-section 5.3 and appendix B.

\subsection{BC fermions in background fields}

We start by giving some useful tools on spinors in the euclidian space time $%
\mathbb{R}^{4}$; then we describe the Dirac equation of the BC
fermions in a generic background field $\mathcal{F}_{\mu \nu }$.

\subsubsection{From $SO\left( 1,3\right) $ to $SO\left( 4\right) $ and then
to $SO\left( 1,4\right) $}

First, recall that the quarks are confined relativistic particles
described by four dimensional space time fields $u_{\alpha }\left(
\mathbf{x}\right) $ and $d_{\alpha }\left( \mathbf{x}\right) $
transforming as complex
4-component Dirac spinors,%
\begin{equation}
\Psi _{\alpha }\left( \mathbf{x}\right) =\left(
\begin{array}{c}
\Psi _{1}\left( \mathbf{x}\right) \\
\Psi _{2}\left( \mathbf{x}\right) \\
\Psi _{3}\left( \mathbf{x}\right) \\
\Psi _{4}\left( \mathbf{x}\right)%
\end{array}%
\right) .  \label{CD}
\end{equation}%
These particles carry several charges, in particular the spin charge of the $%
SO\left( 1,3\right) $ space time symmetry as well as the electric
and color
charges respectively given by the $U_{em}\left( 1\right) $ and the $%
SU_{c}\left( 3\right) $ gauge symmetries. The color charge is
understood in eq(\ref{CD}); since each component $\Psi _{\alpha }$
should be read as $\Psi _{\alpha }^{c}$ with $c=1,2,3$ the color
index; i.e:
\begin{equation}
\Psi _{\alpha }=\left(
\begin{array}{c}
\Psi _{\alpha }^{1} \\
\Psi _{\alpha }^{2} \\
\Psi _{\alpha }^{3}%
\end{array}%
\right) .
\end{equation}%
Recall also that in 4D lattice QCD, one uses euclidian time rather
than the Minkowski one, so the standard $SO\left( 1,3\right) $
Lorentz symmetry gets mapped to $SO\left( 4\right) $ which is
locally isometric to the product
\begin{equation}
SU_{L}\left( 2\right) \times SU_{R}\left( 2\right) .
\end{equation}%
So $SO\left( 1,3\right) $ scalars type $k_{\mu }x^{\mu }=\mathbf{k}.\mathbf{r%
}-k_{4}x^{4}$ get replaced by the $SO\left( 4\right) $ ones namely $\mathbf{k%
}.\mathbf{r}+k_{4}x^{4}$; similarly the $SO\left( 1,3\right) $ semi-norms $%
\mathbf{k}^{2}-k_{4}^{2}=\zeta ^{2}$ transform into positive norms $\mathbf{k%
}^{2}+k_{4}^{2}=\zeta ^{2}$ leading to $k_{4}=\pm \sqrt{\zeta ^{2}-\mathbf{k}%
^{2}}$ with imaginary values in the case where $\zeta =0$. Moreover,
the
four component $SO\left( 4\right) $ Dirac spinor $\Psi _{\alpha }$ (\ref{CD}%
) is essentially made by the direct sum of the two $SU\left(
2\right) $ spinors
\begin{equation}
\Psi _{\alpha }=\psi _{a}\oplus \bar{\xi}_{\dot{a}}\hspace{1cm},\hspace{1cm}%
\Psi =\Psi _{L}\oplus \Psi _{R},
\end{equation}%
where undotted fermions refer to $SU_{L}\left( 2\right) $ and dotted
ones to $SU_{R}\left( 2\right) $. Each one of these Weyl spinors
have two components
as described in the following table,%
\begin{eqnarray}
&&%
\begin{tabular}{l|l|l|l|l}
symmetry groups & \multicolumn{2}{|l}{$SU_{L}\left( 2\right) \times
SU_{R}\left( 2\right) $} & $\simeq SO\left( 4\right) $ & $\ \ \
SU_{c}\left( 3\right) $ \\ \hline representations$\left.
\begin{array}{c}
\\
\\
\\
\\
\end{array}%
\right. $ & $\left( \frac{1}{2},0\right) $ & $\left(
0,\frac{1}{2}\right) $ & $\left( \frac{1}{2},0\right) \oplus \left(
0,\frac{1}{2}\right) $ & $\ \ 3$
\\ \hline
fields$\left.
\begin{array}{c}
\\
\\
\\
\\
\end{array}%
\right. $ & $\ \psi _{a}$ & $\ \ \bar{\xi}_{\dot{a}}$ & $\Psi
_{\alpha }=\left(
\begin{array}{c}
\psi _{a} \\
\bar{\xi}_{\dot{a}}%
\end{array}%
\right) $ & $\left(
\begin{array}{c}
\Psi _{\alpha }^{1} \\
\Psi _{\alpha }^{2} \\
\Psi _{\alpha }^{3}%
\end{array}%
\right) $ \\ \hline
\end{tabular}
\\
&&  \notag
\end{eqnarray}%
The fields $\psi _{a}\left( \mathbf{x}\right) $ and $\bar{\xi}_{\dot{a}%
}\left( \mathbf{x}\right) $ are just the left handed and right
handed quarks
$\Psi _{L}\left( \mathbf{x}\right) $ and $\Psi _{R}\left( \mathbf{x}\right) $%
; they \textrm{carry opposite chiralities under} $\gamma _{5}$ and
live on
the four dimensional Euclidean space time $\mathbb{R}^{4}$ \textrm{%
parameterized by the }local coordinates $\mathbf{x}=\left(
x,y,z,\tau \right) $ with diagonal metric $\delta ^{\mu \nu
}=diag\left( ++++\right) $. These fermions have integer color
charges; but fractional electric ones given by $Q_{u}=\frac{2e}{3}$
and $Q_{d}=-\frac{e}{3}$.\newline \ \ \ \newline
Moreover, in lattice QCD the \emph{BC}- fermions $\tilde{\Psi}\left( \mathbf{%
k}\right) $ living at the Dirac points (\ref{ZM}) have zero energy
as they are zero modes of the Dirac operator. Therefore non zero
energy excitations induced by deformations near the two Dirac points
may be interpreted in terms of the 5-dimensional\emph{\ }direction
$t$ introduced above; so that the space with local coordinates
$\mathbf{x}=\left( x,y,z,\tau \right) $ gets promoted to a (1+4)
space time with coordinates
\begin{equation*}
\left( \mathbf{x};t\right) \equiv \left( x,y,z,\tau ;t\right) .
\end{equation*}%
As such the four component field $\Psi \left( \mathbf{x}\right) $
gets also promoted to $\mathbf{\Psi }\left( \mathbf{x};t\right)
=\mathbf{\Psi }\left( x,y,z,\tau ;t\right) $ whose expansion with
respect to the fifth coordinate t is given by eq(\ref{311}).

\subsubsection{Dirac equation coupled to $\mathcal{F}_{\protect\mu \protect%
\nu }$}

The Dirac equation describing the dynamics of the four component
Dirac fermion $\Psi _{\alpha }^{c}\left( \mathbf{x}\right) $ coupled
to the background vector gauge potential $A_{\mu }\left(
\mathbf{x}\right) $ is given by the following system of four coupled
equations
\begin{equation}
\sum_{\mu =1}^{4}\sum_{\beta =1}^{4}\sum_{d=1}^{3}i\left( \gamma
^{\mu
}\right) _{\alpha }^{\beta }\left[ \delta _{cd}\partial _{\mu }-\frac{i%
\mathrm{g}_{G}}{\mathrm{c}}\left( A_{\mu }^{G}\right) _{cd}\right]
\Psi _{\beta }^{d}=E\Psi _{\alpha }^{c},
\end{equation}%
where $\left( A_{\mu }^{G}\right) _{cd}=\sum_{I}A_{\mu }^{I}\mathcal{T}%
_{cd}^{I}$ with the generators of $G=SU_{c}\left( 3\right) \times
U_{em}\left( 1\right) $ and $\mathrm{g}_{G}$ the gauge coupling
constant. Below, we will restrict to the particular case where
$A_{\mu }^{U_{em}\left(
1\right) }\neq 0$ and $A_{\mu }^{SU_{c}\left( 3\right) }=0$; the case with $%
\mathcal{F}_{\mu \nu }^{SU_{c}\left( 3\right) }\neq 0$ will be
discussed in conclusion section. In the case $A_{\mu }^{U_{em}\left(
1\right) }\neq 0$
and $A_{\mu }^{SU_{c}\left( 3\right) }=0$, the above equation reduces to,%
\begin{equation}
i\gamma ^{\mu }\left( \partial _{\mu }-i\frac{Q}{c}A_{\mu }\right)
\Psi =E\Psi ,  \label{DIR}
\end{equation}%
where now $A_{\mu }$ stands for the electromagnetic gauge potential
$A_{\mu }^{U_{em}\left( 1\right) }$. Notice also the two following
things: \newline First, by help of the constraint equation
$\frac{\partial \mathcal{F}_{\mu \nu }}{\partial x^{\rho }}=0$
expressing the fact that $\mathcal{F}_{\mu \nu }$ is a constant, it
is not difficult to check that the electromagnetic
potential vector $A_{\mu }$ is linear in space time coordinate positions $%
x^{\mu }$ as follows,
\begin{equation}
\begin{tabular}{ll}
$A_{\mu }=\frac{1}{2}\mathcal{F}_{\mu \nu }x^{\nu },$ & $\partial
^{\mu
}A_{\mu }=\frac{\partial A_{\mu }}{\partial x_{\mu }}=\frac{1}{2}\mathcal{F}%
_{\mu \nu }\delta ^{\mu \nu }=\frac{1}{2}Tr\left( \mathcal{F}\right) =0,$%
\end{tabular}
\label{AM}
\end{equation}%
where we have dropped out the irrelevant integration constants.
Generally speaking, the antisymmetric tensor $\mathcal{F}_{\mu \nu
}=\partial _{\mu }A_{\nu }-\partial _{\nu }A_{\mu }$ capture six
real degrees of freedom, three components for the "magnetic" field
$\mathcal{B}$ and three components of the "electric" field
$\mathcal{E},$
\begin{equation}
\mathcal{F}_{\mu \nu }=\left(
\begin{array}{cccc}
0 & +\mathcal{B}_{3} & -\mathcal{B}_{2} & +\mathcal{E}_{1} \\
-\mathcal{B}_{3} & 0 & +\mathcal{B}_{1} & +\mathcal{E}_{2} \\
+\mathcal{B}_{2} & -\mathcal{B}_{1} & 0 & +\mathcal{E}_{3} \\
-\mathcal{E}_{1} & -\mathcal{E}_{2} & -\mathcal{E}_{3} & 0%
\end{array}%
\right) ,\qquad \det \mathcal{F}_{\mu \nu }=(\mathcal{\vec{E}}.\mathcal{\vec{%
B})}^{2}.  \label{GEN}
\end{equation}%
From the 5-dimensional view, this matrix should be reads as a
4$\times $4
submatrix of the following 5$\times $5 antisymmetric one,%
\begin{equation}
\mathcal{F}_{MN}=\left(
\begin{array}{cc}
\mathcal{F}_{\mu \nu } & \mathcal{F}_{\mu 5} \\
\mathcal{F}_{5\nu } & 0%
\end{array}%
\right) ,
\end{equation}%
where now the $\mathcal{F}_{\mu 5}$'s are the \emph{4} components of
the electric field in $5D$; and $\mathcal{F}_{\mu \nu }$ the
\emph{6} components of the magnetic tensor. Below, we will focuss on
particular situations where $\mathcal{F}_{\mu \nu }$ has two degrees
of freedom as in eq(\ref{FF}). This corresponds to choose the four
components of the electromagnetic gauge
potential $A_{\mu }$ as follows:%
\begin{equation}
\begin{tabular}{llll}
$A_{1}=\frac{\mathcal{B}}{2}y,$ & $A_{2}=\frac{-\mathcal{B}}{2}x,$ & $A_{3}=%
\frac{\mathcal{E}}{2}\tau ,$ & $A_{4}=-\frac{\mathcal{E}}{2}z,$%
\end{tabular}
\label{MA}
\end{equation}%
leading to the following constant electromagnetic field strength,
\begin{equation}
\mathcal{F}_{\mu \nu }=\left(
\begin{array}{cccc}
0 & -\mathcal{B} & 0 & 0 \\
+\mathcal{B} & 0 & 0 & 0 \\
0 & 0 & 0 & -\mathcal{E} \\
0 & 0 & +\mathcal{E} & 0%
\end{array}%
\right) ,  \label{FAC}
\end{equation}%
with $\det \mathcal{F}_{\mu \nu }=\mathcal{B}^{2}\times
\mathcal{E}^{2}$. In this relation, $\mathcal{B}$ and $\mathcal{E}$
are a priori arbitrary real numbers; but to have a quantum Hall
effect, they have to be large enough so
that interacting energy is much greater with respect to kinetic energy.%
\newline
The second comment concerns the four $\gamma ^{\mu }$'s of
(\ref{DIR}); these are the \emph{4D} euclidian $4\times 4$ Dirac
matrices obeying the
usual \emph{4D} Clifford algebra,%
\begin{equation}
\begin{tabular}{ll}
$\gamma ^{\mu }\gamma ^{\nu }+\gamma ^{\nu }\gamma ^{\mu }=2\delta
^{\mu \nu
}I_{4}$ & ,%
\end{tabular}
\label{G}
\end{equation}%
with $I_{4}$ the $4\times 4$ identity matrix. Notice that the
$SO\left( 4\right) \simeq $ $SU_{L}\left( 2\right) \times
SU_{R}\left( 2\right) $ has 6 generators, 3 of them generate
$SU_{L}\left( 2\right) $ and the three others generate $SU_{R}\left(
2\right) $. In the spinor representation, these are given by the
Pauli matrices $\tau ^{i}$ and $\sigma ^{i}$ with
generic $2\times 2$ matrix representation is as follows,%
\begin{equation}
\begin{tabular}{lll}
$\varrho ^{1}=\left(
\begin{array}{cc}
0 & 1 \\
1 & 0%
\end{array}%
\right) ,$ & $\varrho ^{2}=\left(
\begin{array}{cc}
0 & -i \\
i & 0%
\end{array}%
\right) ,$ & $\varrho ^{3}=\left(
\begin{array}{cc}
1 & 0 \\
0 & -1%
\end{array}%
\right) ,$%
\end{tabular}%
\end{equation}%
with $\varrho ^{i}$ standing for both $\tau ^{i}$ and $\sigma ^{i}$.
The
matrices $\varrho ^{1}$, $\varrho ^{2}$ satisfy the 2D Clifford algebra $%
\varrho ^{i}\varrho ^{j}+\varrho ^{j}\varrho ^{i}=2\delta
^{ij}I_{2}$ and
the $SU\left( 2\right) $ symmetry bracket $\left[ \varrho ^{1},\varrho ^{2}%
\right] =2i\varrho ^{3}$. Notice moreover that the euclidian $\gamma
^{\mu }$
matrices can be realized in terms of tensor products of $\tau ^{i}$ and $%
\sigma ^{i}$ as follows,%
\begin{equation}
\begin{tabular}{llllllll}
$\gamma ^{i}=\tau ^{2}\otimes \sigma ^{i}$ & , & $\gamma ^{4}=\tau
^{1}\otimes \sigma ^{4}$ & , & $\gamma ^{5}=\tau ^{3}\otimes \sigma
^{4}$ & , & $\Upsilon ^{0}=\tau ^{4}\otimes \sigma ^{4}$ &
\end{tabular}
\label{GAM}
\end{equation}%
with $\tau ^{4}$, $\sigma ^{4}=I_{2}\equiv I$ and $\gamma
^{5}=\gamma ^{1}\gamma ^{2}\gamma ^{3}\gamma ^{4}$. More explicitly,
we have
\begin{equation}
\begin{tabular}{llll}
$\gamma ^{k}{\small =}\left(
\begin{array}{cc}
{\small 0} & {\small -i\sigma }^{k} \\
{\small i\sigma }^{k} & {\small 0}%
\end{array}%
\right) ,$ & $\gamma ^{4}{\small =}\left(
\begin{array}{cc}
{\small 0} & {\small I} \\
{\small I} & {\small 0}%
\end{array}%
\right) $, & $\gamma ^{5}{\small =}\left(
\begin{array}{cc}
{\small I} & {\small 0} \\
{\small 0} & {\small -I}%
\end{array}%
\right) $ & $\Upsilon ^{0}{\small =}\left(
\begin{array}{cc}
{\small I} & {\small 0} \\
{\small 0} & {\small I}%
\end{array}%
\right) $. \\
&  &  &
\end{tabular}
\label{GA}
\end{equation}%
The anticommutator of the $\gamma ^{\mu }$ matrices verify the
\emph{4D}
Clifford algebra (\ref{G}) and their commutators $\gamma ^{\left[ \mu \nu %
\right] }$ give precisely the \emph{6} generators of the spinorial
representation of the $SO\left( 4\right) $ symmetry,%
\begin{equation}
\begin{tabular}{lll}
$\gamma ^{i}\gamma ^{j}-\gamma ^{j}\gamma ^{i}$ & $=2i\varepsilon
^{ijk}\left( \tau ^{2}\otimes \sigma ^{k}\right) $ & , \\
$\gamma ^{4}\gamma ^{i}-\gamma ^{i}\gamma ^{4}$ & $=2i\varepsilon
^{123}\left( \tau ^{3}\otimes \sigma ^{i}\right) $ & ,%
\end{tabular}%
\end{equation}%
where $\varepsilon ^{ijk}$ is the completely antisymmetric \emph{3D}
Levi-Civita tensor. \newline Now, we turn to determine the spectrum
of BC fermions in background fields by solving eq(\ref{DIR}); but
before that it is interesting to study the properties of the non
commutative algebra of the covariant derivatives; then turn back to
compute the spectrum.

\section{The Lie algebra of the gauge covariant derivatives}

As we will see, the algebra of the $D_{\mu }$ covariant derivatives
in constant background fields has much to do with Heisenberg algebra
of quantum harmonic oscillators. Because of the tensor structure of
$\mathcal{F}_{\mu \nu }$, we distinguish two basic cases: first the
case of two uncoupled quantum oscillators associated with the choice
(\ref{FAC}) and $\det \mathcal{F}_{\mu \nu }\neq 0$. Then we
consider the general case (\ref{GEN})
with $\det \mathcal{F}_{\mu \nu }=(\mathcal{\vec{E}}.\mathcal{\vec{B})}%
^{2}\neq 0$ describing two coupled quantum harmonic oscillators.

\subsection{uncoupled case: $\ \mathcal{F}_{\protect\mu \protect\nu }=%
\mathcal{B}\protect\varepsilon _{\protect\mu \protect\nu 34}+\mathcal{E}%
\protect\varepsilon _{12\protect\mu \protect\nu }$}

Generally, the four gauge covariant derivatives $D_{1},$ $D_{2},$ $D_{3},$ $%
D_{4}$ satisfy the generic commutation relations
\begin{equation}
\left[ D_{\mu },D_{\nu }\right] =-i\frac{Q}{c}\mathcal{F}_{\mu \nu
},
\end{equation}%
but because of the choice (\ref{FAC}) of the background fields, they
reduce
to,%
\begin{equation}
\begin{tabular}{llll}
$\left[ D_{1},D_{2}\right] =i\frac{Q\mathcal{B}}{c}$ & $,$ & $\left[
D_{3},D_{4}\right] =i\frac{Q\mathcal{E}}{c}$ & $,$%
\end{tabular}
\label{CR}
\end{equation}%
and all others vanishing. These are very special relations; first
because their right hand sides are non zero constant numbers showing
amongst others that the large limits of $\mathcal{B}$ and
$\mathcal{E}$ induce a non
commutative geometry in the euclidian space,%
\begin{equation}
\left[ x^{\mu },x^{\nu }\right] =\frac{4ic}{Q}\mathcal{G}^{\mu \nu }
\end{equation}%
with $\mathcal{G}^{\mu \nu }$ given by%
\begin{equation}
\mathcal{G}^{\mu \nu
}=\frac{1}{\mathcal{\vec{E}}.\mathcal{\vec{B}}}\left(
\begin{array}{cccc}
0 & -\mathcal{E}_{3} & \mathcal{E}_{2} & -\mathcal{B}_{1} \\
\mathcal{E}_{3} & 0 & -\mathcal{E}_{1} & -\mathcal{B}_{2} \\
-\mathcal{E}_{2} & \mathcal{E}_{1} & 0 & -\mathcal{B}_{3} \\
\mathcal{B}_{1} & \mathcal{B}_{2} & \mathcal{B}_{3} & 0%
\end{array}%
\right)  \label{GG}
\end{equation}%
Second, the four covariant derivatives organize in $2+2$ capturing a
complex
structure as follows,%
\begin{eqnarray}
&&%
\begin{tabular}{llll}
$D_{1}-iD_{2}=\frac{2\partial }{\partial
u}+\frac{Q\mathcal{B}}{2c}\bar{u}$
& $,$ & $D_{1}+iD_{2}=\frac{2\partial }{\partial \bar{u}}-\frac{Q\mathcal{B}%
}{2c}u$ & $,$ \\
&  &  &  \\
$D_{3}-iD_{4}=\frac{2\partial }{\partial
v}+\frac{Q\mathcal{E}}{2c}\bar{v}$
& $,$ & $D_{3}+iD_{4}=\frac{2\partial }{\partial \bar{v}}-\frac{Q\mathcal{E}%
}{2c}v$ & $,$%
\end{tabular}
\label{RC} \\
&&  \notag
\end{eqnarray}%
together with similar relations for their adjoint conjugates. In getting eq(%
\ref{RC}), we have used the explicit expressions $D_{1}=\partial _{1}-i\frac{%
Q\mathcal{B}}{2c}y$, $D_{2}=\partial _{2}+i\frac{Q\mathcal{B}}{2c}x$
together with similar relations for $D_{3}$, $D_{4}$; and we have set%
\begin{equation}
\begin{tabular}{llll}
$u=x+iy$ & $,$ & $\frac{\partial }{\partial u}=\frac{1}{2}\left( \frac{%
\partial }{\partial x}-i\frac{\partial }{\partial y}\right) $ & , \\
$v=z+i\tau $ & , & $\frac{\partial }{\partial v}=\frac{1}{2}\left( \frac{%
\partial }{\partial z}-i\frac{\partial }{\partial \tau }\right) $ & .%
\end{tabular}
\label{CV}
\end{equation}%
To study the representations of eqs(\ref{CR}), one has to specify
the sign
of $\frac{Q\mathcal{B}}{c}$ and $\frac{Q\mathcal{E}}{c}$. Assuming $\frac{Q%
\mathcal{B}}{c}>0,$ $\frac{Q\mathcal{E}}{c}>0$ for simplicity and
which means that $Q$, $\mathcal{B}$, $\mathcal{E}$ have the same
sign; and setting
\begin{equation}
\begin{tabular}{llll}
$i\left( D_{1}-iD_{2}\right) =A\sqrt{\frac{2Q\mathcal{B}}{c}}$ & , & $%
i\left( D_{1}+iD_{2}\right) =A^{\dagger
}\sqrt{\frac{2Q\mathcal{B}}{c}}$ & ,
\\
$i\left( D_{3}-iD_{4}\right) =C\sqrt{\frac{2Q\mathcal{E}}{c}}$ & , & $%
i\left( D_{3}+iD_{4}\right) =C^{\dagger }\sqrt{\frac{2Q\mathcal{E}}{c}}$ & ,%
\end{tabular}
\label{AAC}
\end{equation}%
the commutation relations (\ref{CR}) read also as
\begin{equation}
\begin{tabular}{llll}
$\left[ A,A^{\dagger }\right] =I$ & , & $\left[ C,C^{\dagger
}\right] =I$ & ,
\\
$\left[ A,C\right] =0$ & , & $\left[ A,C^{\dagger }\right] =0$ & ;%
\end{tabular}
\label{COM}
\end{equation}%
These relations (\ref{COM}) show that the Dirac fermion in the
background field $\mathcal{F}_{\mu \nu }$ (\ref{FAC}) describe a
priori two quantum harmonic oscillators with oscillation frequencies
\begin{equation}
\varpi =\sqrt{\frac{2Q\mathcal{B}}{c}}\qquad ,\qquad \varpi ^{\prime }=\sqrt{%
\frac{2Q\mathcal{E}}{c}}.  \label{OME}
\end{equation}%
The operators $A^{\dagger }A$ and $C^{\dagger }C$ giving the number
of energy excitations in $\varpi $ and $\varpi ^{\prime }$ units
respectively read in terms of the gauge covariant derivatives as
follows
\begin{equation}
\begin{tabular}{llll}
$\frac{2Q\mathcal{B}}{c}A^{\dagger }A$ & = & $\left( D_{1}\right)
^{2}+\left( D_{2}\right) ^{2}+i\left[ D_{1},D_{2}\right] $ &  \\
& = & $\left( D_{1}\right) ^{2}+\left( D_{2}\right) ^{2}-\frac{Q\mathcal{B}}{%
c}$ & ,%
\end{tabular}%
\end{equation}%
and similarly%
\begin{equation}
\begin{tabular}{llll}
$\frac{2Q\mathcal{E}}{c}C^{\dagger }C$ & = & $\left( D_{3}\right)
^{2}+\left( D_{4}\right) ^{2}+i\left[ D_{3},D_{4}\right] $ &  \\
& = & $\left( D_{3}\right) ^{2}+\left( D_{4}\right) ^{2}-\frac{Q\mathcal{E}}{%
c}$ & .%
\end{tabular}%
\end{equation}%
Using the space time variables $\left( x,y,z,\tau \right) $ and
their derivatives $\partial _{\mu }$, these operators read more
explicitly like
\begin{equation}
\begin{tabular}{ll}
$\left( D_{1}\right) ^{2}+\left( D_{2}\right) ^{2}$ & $=-\frac{\partial ^{2}%
}{\partial x^{2}}-\frac{\partial ^{2}}{\partial y^{2}}+\left\vert \frac{Q%
\mathcal{B}}{2c}\right\vert ^{2}\left( x^{2}+y^{2}\right) +\left\vert \frac{Q%
\mathcal{B}}{2c}\right\vert \mathcal{L}_{xy},$ \\
&  \\
& $=-\frac{4\partial ^{2}}{\partial u\partial \bar{u}}+\left\vert \frac{Q%
\mathcal{B}}{2c}\right\vert ^{2}u\bar{u}+\left\vert \frac{Q\mathcal{B}}{2c}%
\right\vert \mathcal{L}_{u\bar{u}},$%
\end{tabular}
\label{11}
\end{equation}%
and%
\begin{eqnarray}
&&%
\begin{tabular}{ll}
$\left( D_{3}\right) ^{2}+\left( D_{4}\right) ^{2}$ & $=-\frac{\partial ^{2}%
}{\partial z^{2}}-\frac{\partial ^{2}}{\partial \tau ^{2}}+\left\vert \frac{Q%
\mathcal{E}}{2c}\right\vert ^{2}\left( z^{2}+\tau ^{2}\right)
+\left\vert
\frac{Q\mathcal{E}}{2c}\right\vert \mathcal{L}_{z\tau },$ \\
&  \\
& $=-\frac{4\partial ^{2}}{\partial v\partial \bar{v}}+\left\vert \frac{Q%
\mathcal{E}}{2c}\right\vert ^{2}\left( v\bar{v}\right) +\left\vert \frac{Q%
\mathcal{E}}{2c}\right\vert \mathcal{L}_{v\bar{v}},$%
\end{tabular}
\label{12} \\
&&  \notag
\end{eqnarray}%
where one recognizes the harmonic potential energy induced by the
electromagnetic flux and the angular momentum operator components $\mathcal{L%
}_{xy}$ and $\mathcal{L}_{z\tau }$
\begin{equation}
\begin{tabular}{llll}
$\mathcal{L}_{xy}=-i\left( x\frac{\partial }{\partial y}-y\frac{\partial }{%
\partial x}\right) $ & , & $\mathcal{L}_{z\tau }=-i\left( z\frac{\partial }{%
\partial \tau }-\tau \frac{\partial }{\partial z}\right) $ & ,%
\end{tabular}
\label{LL}
\end{equation}%
describing respectively rotations around of the oscillations of the
particles the $x$-$y$ and $z$-$\tau $ planes of the 4D space time.
Notice that the hermitian operators $\mathcal{L}_{xy}$ and
$\mathcal{L}_{z\tau }$\ are nothing but the charge operators of the
abelian $U_{L}\left( 1\right) $ and $U_{R}\left( 1\right) $
sub-symmetries of the $SU_{L}\left( 2\right) $ and $SU_{R}\left(
2\right) $\ respectively. Indeed, using the complex variables
(\ref{CV}) and putting back into (\ref{LL}), we find,
\begin{equation}
\begin{tabular}{llll}
$\mathcal{L}_{xy}=\mathcal{L}_{u\bar{u}}=u\frac{\partial }{\partial u}-\bar{u%
}\frac{\partial }{\partial \bar{u}}$ & , & $\mathcal{L}_{z\tau }=\mathcal{L}%
_{v\bar{v}}=v\frac{\partial }{\partial v}-\bar{v}\frac{\partial
}{\partial
\bar{v}}$ & ,%
\end{tabular}%
\end{equation}%
acting on the monomials
$u^{n_{1}}\bar{u}^{n_{2}}v^{m_{1}}\bar{v}^{m_{2}}$\
as follows%
\begin{eqnarray}
&&%
\begin{tabular}{llll}
$\left[ \mathcal{L}_{xy},u^{n_{1}}\bar{u}^{n_{2}}v^{m_{1}}\bar{v}^{m_{2}}%
\right] $ & $=$ & $\left( n_{1}-n_{2}\right) u^{n_{1}}\bar{u}%
^{n_{2}}v^{m_{1}}\bar{v}^{m_{2}}$ & , \\
&  &  &  \\
$\left[ \mathcal{L}_{z\tau },u^{n_{1}}\bar{u}^{n_{2}}v^{m_{1}}\bar{v}^{m_{2}}%
\right] $ & $=$ & $\left( m_{1}-m_{2}\right) u^{n_{1}}\bar{u}%
^{n_{2}}v^{m_{1}}\bar{v}^{m_{2}}$ & .%
\end{tabular}
\\
&&  \notag
\end{eqnarray}%
Notice that for the particular cases $n_{1}=n_{2}$ and
$m_{1}=m_{2}$, there is no $U_{L}\left( 1\right) $ nor $U_{R}\left(
1\right) $ charges. So wave functions $\Psi _{\alpha }=\Psi _{\alpha
}\left( u,v,\bar{u},\bar{v}\right) $
that depend only on $u\bar{u}$ and $v\bar{v}$ have no angular momenta.%
\newline
Moreover, using the expression\textrm{\ }of the matrices $\gamma
^{\mu }$, we can write this matrix operator $\left( H_{BC}\right)
_{\alpha \beta }=i\gamma _{\alpha \beta }^{\mu }D_{\mu }$ in terms
of the creations operators $A^{\dagger },$ $C^{\dagger }$ and the
annihilation ones $A,$ $C$
as follows:%
\begin{eqnarray}
\left( H_{BC}\right) _{\alpha \beta } &=&\frac{1}{i}\left(
\begin{array}{cccc}
0 & 0 & -\varpi ^{\prime }C^{\dagger } & -\varpi A \\
0 & 0 & -\varpi A^{\dagger } & \varpi ^{\prime }C \\
\varpi ^{\prime }C & \varpi A & 0 & 0 \\
\varpi A^{\dagger } & -\varpi ^{\prime }C^{\dagger } & 0 & 0%
\end{array}%
\right) ,  \label{DC} \\
&&  \notag
\end{eqnarray}%
with $H_{BC}^{\dagger }=H_{BC}$. Moreover, using the commutation properties $%
CA^{\dagger }=A^{\dagger }C$ and $CA=AC$ and their adjoint
conjugates
following from the choice (\ref{FAC}), the squared \emph{BC} hamiltonian $%
H_{BC}^{2}$ reads as follows,%
\begin{eqnarray*}
&&\left(
\begin{array}{cccc}
\varpi ^{2}AA^{\dagger }+\varpi ^{\prime 2}C^{\dagger }C & 0 & 0 & 0 \\
0 & \varpi ^{2}A^{\dagger }A+\varpi ^{\prime 2}CC^{\dagger } & 0 & 0 \\
0 & 0 & \varpi ^{2}AA^{\dagger }+\varpi ^{\prime 2}CC^{\dagger } & 0 \\
0 & 0 & 0 & \varpi ^{2}A^{\dagger }A+\varpi ^{\prime 2}C^{\dagger }C%
\end{array}%
\right) \\
&&
\end{eqnarray*}%
which allow to determine spectrum of the Bori\c{c}i-Creutz spectrum;
this will be done after studying the coupled case.

\subsection{coupled case: $\ \mathcal{F}_{\protect\mu \protect\nu }=\mathcal{%
B}^{\left[ ij\right] }\protect\varepsilon _{\protect\mu \protect\nu ij}+%
\mathcal{E}^{\left[ 4i\right] }\protect\varepsilon _{4i\protect\mu \protect%
\nu }$}

In the case where the \emph{6} components $\mathcal{B}^{\left[
ij\right] }$ and $\mathcal{E}^{\left[ 4i\right] }$, ($i,j=1,2,3$) of
the background field $\mathcal{F}_{\mu \nu }$ are non zero as in
(\ref{GEN}), the previous two
Heisenberg algebras (\ref{COM}) get coupled. Indeed, using the relation $%
\left[ D_{\mu },D_{\nu }\right] =-i\frac{Q}{c}\mathcal{F}_{\mu \nu }$ and eq(%
\ref{AAC}), we find in addition to
\begin{equation}
\begin{tabular}{llll}
$\left[ A,A^{\dagger }\right] =I$ & , & $\left[ C,C^{\dagger }\right] =I$ & ,%
\end{tabular}%
\end{equation}%
the following couplings
\begin{equation}
\begin{tabular}{llll}
$\left[ A,C\right] $ & $=\lambda $, & $\left[ A^{\dagger },C^{\dagger }%
\right] $ & $=-\bar{\lambda}$, \\
$\left[ A,C^{\dagger }\right] $ & $=\zeta $, & $\left[ A^{\dagger
},C\right]
$ & $=-\bar{\zeta}$,%
\end{tabular}%
\end{equation}%
where we have set,%
\begin{eqnarray}
&&%
\begin{tabular}{lll}
$\lambda =$ & $\frac{1}{2\sqrt{\mathcal{BE}}}\left[ \left( \mathcal{F}_{14}+%
\mathcal{F}_{23}\right) +i\left( \mathcal{F}_{13}-\mathcal{F}_{24}\right) %
\right] $ & , \\
&  &  \\
$\zeta =$ & $\frac{1}{2\sqrt{\mathcal{BE}}}\left[ \left( \mathcal{F}_{23}-%
\mathcal{F}_{14}\right) +i\left( \mathcal{F}_{13}+\mathcal{F}_{24}\right) %
\right] $ & .%
\end{tabular}
\label{LAM} \\
&&  \notag
\end{eqnarray}%
Using the \emph{BC} hamiltonian (\ref{DC}), we find that its square $%
H_{BC}^{2}$ is given by,
\begin{eqnarray*}
&&\left(
\begin{array}{cccc}
\varpi ^{2}AA^{\dagger }+\varpi ^{\prime 2}C^{\dagger }C & -\varpi
\varpi
^{\prime }\zeta & 0 & 0 \\
-\varpi \varpi ^{\prime }\bar{\zeta} & \varpi ^{2}A^{\dagger
}A+\varpi
^{\prime 2}CC^{\dagger } & 0 & 0 \\
0 & 0 & \varpi ^{2}AA^{\dagger }+\varpi ^{\prime 2}CC^{\dagger } &
-\varpi
\varpi ^{\prime }\lambda \\
0 & 0 & -\varpi \varpi ^{\prime }\bar{\lambda}\  & \varpi
^{2}A^{\dagger
}A+\varpi ^{\prime 2}C^{\dagger }C%
\end{array}%
\right) \\
&&
\end{eqnarray*}%
and is no longer diagonal. Moreover, the operator numbers
$A^{\dagger }A$ and $C^{\dagger }C$\ do no longer commute; a
difficulty that may be overcome by treading $\lambda $\ and $\zeta $
as small perturbation parameters.

\section{More on AQHE in BC model}

First we compute the spectrum of BC fermions in the constant
background fields; then we determine the filling factor $\nu _{BC}$
of the associated quantum Hall effect. After that, we study the link
with spectral flow method of \textrm{\cite{4K,2BB}.}

\subsection{Spectrum of \emph{BC} Hamiltonian $H_{BC}$}

Starting from the BC fermion's equation $H_{BC}\left\vert \Psi
\right\rangle
=E\left\vert \Psi _{E}\right\rangle ,$ then substituting the matrices $%
\gamma ^{\mu }$ by their expressions (\ref{GA}) and the Dirac
fermion $\Psi
_{\alpha }$ by its two Weyl spinors $\left( \psi _{a},\bar{\xi}_{\dot{a}%
}\right) $, we get%
\begin{eqnarray}
\left(
\begin{array}{cc}
0 & \mathcal{O} \\
\mathcal{O}^{\dagger } & 0%
\end{array}%
\right) \left(
\begin{array}{c}
\psi  \\
\bar{\xi}%
\end{array}%
\right)  &=&E\left(
\begin{array}{c}
\psi  \\
\bar{\xi}%
\end{array}%
\right) ,  \label{4D} \\
&&  \notag
\end{eqnarray}%
where the $\mathcal{O}$ and $\mathcal{O}^{\dagger }$ operators are given by%
\begin{equation}
\begin{tabular}{llll}
$\mathcal{O}=i\left(
\begin{array}{cc}
-\varpi ^{\prime }C^{\dagger } & -\varpi A \\
-\varpi A^{\dagger } & \varpi ^{\prime }C%
\end{array}%
\right) $ & $,$ & $\mathcal{O}^{\dagger }=i\left(
\begin{array}{cc}
\varpi ^{\prime }C & \varpi A \\
\varpi A^{\dagger } & -\varpi ^{\prime }C^{\dagger }%
\end{array}%
\right) $ & $.$%
\end{tabular}%
\end{equation}%
These operators can be also written in the following condensed form
\begin{eqnarray}
\mathcal{O} &=&\sigma ^{4}D_{4}+i\dsum\limits_{k=1}^{3}\sigma
^{k}D,\qquad \mathcal{O}^{\dagger }=\sigma
^{4}D_{4}-i\dsum\limits_{k=1}^{3}\sigma ^{k}D,
\\
&&  \notag
\end{eqnarray}%
which let understand that solutions may be as well obtained by using
quaternions. To deal with eqs(\ref{4D}), we will use an explicit
method relying on acting on its both sides of by $H_{BC}$. We get
the following set of equations
\begin{equation}
\begin{tabular}{lll}
$\mathcal{O\mathcal{O}}^{\dagger }\psi $ & $=E^{2}\psi $ & , \\
$\mathcal{O}^{\dagger }\mathcal{O}\bar{\xi}$ & $=E^{2}\bar{\xi}$ & ,%
\end{tabular}
\label{PS}
\end{equation}%
showing that the two Weyl spinors $\psi _{a}$ and
$\bar{\xi}_{\dot{a}}$ can be treated separately; thank to the choice
(\ref{FAC}). This property was also expected as there is no mass
term that couples the left and right components of the Dirac
fermion. To solve (\ref{PS}), we use the following
expression of $\mathcal{OO}^{\dagger }$ and $\mathcal{O}^{\dagger }\mathcal{O%
}$,%
\begin{eqnarray}
\mathcal{OO}^{\dagger } &=&\left(
\begin{array}{cc}
\varpi ^{\prime 2}C^{\dagger }C+\varpi ^{2}\left[ A^{\dagger
}A+1\right]  & 0
\\
0 & \varpi ^{\prime 2}\left[ C^{\dagger }C+1\right] +\varpi ^{2}A^{\dagger }A%
\end{array}%
\right) ,  \label{21} \\
&&  \notag
\end{eqnarray}%
and%
\begin{eqnarray}
\mathcal{O}^{\dagger }\mathcal{O} &=&\left(
\begin{array}{cc}
\varpi ^{\prime 2}\left[ C^{\dagger }C+1\right] +\varpi ^{2}\left[
A^{\dagger }A+1\right]  & 0 \\
0 & \varpi ^{\prime 2}C^{\dagger }C+\varpi ^{2}A^{\dagger }A%
\end{array}%
\right) .  \label{22} \\
&&  \notag
\end{eqnarray}%
Notice that these matrix operators don't commute; they obey the
commutation
relation%
\begin{eqnarray}
\mathcal{OO}^{\dagger }-\mathcal{O}^{\dagger }\mathcal{O} &=&-\varpi
^{\prime 2}\left(
\begin{array}{cc}
1 & 0 \\
0 & -1%
\end{array}%
\right) . \\
&&  \notag
\end{eqnarray}%
Then putting the expression of $\mathcal{OO}^{\dagger }$ and $\mathcal{O}%
^{\dagger }\mathcal{O}$ back into eq(\ref{PS}), we obtain the
following system of \emph{4} equations which allow to determine the
\emph{4} component
of the Dirac spinor%
\begin{equation}
\begin{tabular}{lll}
$\ \ \ \ \ \ \ \left[ \varpi ^{\prime 2}C^{\dagger }C+\varpi
^{2}\left[
A^{\dagger }A+1\right] \right] \psi _{1}$ & $=E^{2}\psi _{1}$ & , \\
&  &  \\
$\ \ \ \ \ \ \ \left[ \varpi ^{\prime 2}\left[ C^{\dagger
}C+1\right]
+\varpi ^{2}A^{\dagger }A\right] \psi _{2}$ & $=E^{2}\psi _{2}$ & , \\
&  &  \\
$\left[ \varpi ^{\prime 2}\left[ C^{\dagger }C+1\right] +\varpi
^{2}\left[
A^{\dagger }A+1\right] \right] \bar{\xi}_{1}$ & $=E^{2}\bar{\xi}_{1}$ & , \\
&  &  \\
$\ \ \ \ \ \ \ \ \ \ \ \ \ \ \ \ \ \left[ \varpi ^{\prime
2}C^{\dagger }C+\varpi ^{2}A^{\dagger }A\right] \bar{\xi}_{2}$ &
$=E^{2}\bar{\xi}_{2}$ & .
\\
&  &
\end{tabular}%
\end{equation}%
To solve these relations, it is useful to introduce the wave function basis $%
\left\{ \left\vert \vartheta _{n,m}\right\rangle \right\} $ of two
uncoupled quantum harmonic oscillators. These wave functions are
factorized as
\begin{equation}
\mathbf{\Lambda }_{n,m}\left( u,\bar{u},v,\bar{v}\right) =\theta
_{n}\left( u,\bar{u}\right) \times \theta _{m}^{\prime }\left(
v,\bar{v}\right) ,\qquad n,m\geq 0,
\end{equation}%
with the constraint relations%
\begin{equation}
\begin{tabular}{ll}
$A\times \theta _{0}\left( u,\bar{u}\right) =\left( \frac{2\partial }{%
\partial u}+\frac{Q\mathcal{B}}{2c}\bar{u}\right) \theta _{0}\left( u,\bar{u}%
\right) =0$ &  \\
&  \\
$C\times \theta _{0}^{\prime }\left( v,\bar{v}\right) =\left( \frac{%
2\partial }{\partial v}+\frac{Q\mathcal{B}}{2c}\bar{v}\right) \theta
_{0}\left( v,\bar{v}\right) =0$ &
\end{tabular}%
\end{equation}%
solved as follows%
\begin{equation}
\begin{tabular}{llll}
$\theta _{0}\left( u,\bar{u}\right) =\mathcal{N}_{0}\left(
\bar{u}\right)
e^{-\frac{Q\mathcal{B}}{4c}u\bar{u}}$ & , & $\theta _{0}^{\prime }\left( v,%
\bar{v}\right) =\mathcal{N}_{0}^{\prime }\left( \bar{v}\right) e^{-\frac{Q%
\mathcal{E}}{4c}v\bar{v}}$ & ,%
\end{tabular}
\label{NN}
\end{equation}%
where the complex functions $\mathcal{N}_{0}\left( \bar{u}\right) $ and $%
\mathcal{N}_{0}^{\prime }\left( \bar{v}\right) $ obey the
holomorphic
conditions $\frac{\partial \mathcal{N}_{0}\left( \bar{u}\right) }{\partial u}%
=0$ and $\frac{\partial \mathcal{N}_{0}^{\prime }\left( \bar{v}\right) }{%
\partial v}=0$; and are generally fixed by the charge of the angular
momentum operators and the normalization condition of the wave
functions. The others waves functions $\theta _{n}\left(
u,\bar{u}\right) $ and $\theta _{m}^{\prime }\left( v,\bar{v}\right)
$ are obtained as usual by applying the creation operators.
\begin{equation}
\begin{tabular}{llll}
$\theta _{n}\left( u,\bar{u}\right) $ & $\simeq $ & $\left(
\frac{2\partial
}{\partial \bar{u}}-\frac{Q\mathcal{B}}{2c}u\right) ^{n}\theta _{0}\left( u,%
\bar{u}\right) $ & , \\
&  &  &  \\
$\theta _{n}^{\prime }\left( v,\bar{v}\right) $ & $\simeq $ & $\left( \frac{%
2\partial }{\partial \bar{v}}-\frac{Q\mathcal{E}}{2c}v\right)
^{n}\theta
_{0}^{\prime }\left( v,\bar{v}\right) $ & . \\
&  &  &
\end{tabular}
\label{W}
\end{equation}%
Because of the algebraic properties $\left[ A^{\dagger }A,\left(
A^{\dagger }\right) ^{n}\right] =n\left( A^{\dagger }\right) ^{n}$
and $\left[ C^{\dagger }C,\left( C^{\dagger }\right) ^{m}\right]
=m\left( C^{\dagger }\right) ^{m}$, the $\left\vert \mathbf{\Lambda
}_{n,m}\right\rangle $ states obey amongst others the following
properties
\begin{equation}
\begin{tabular}{lll}
$\ \ \ \ \ A^{\dagger }A\left\vert \mathbf{\Lambda }_{n,m}\right\rangle $ & $%
=n\left\vert \mathbf{\Lambda }_{n,m}\right\rangle $ & , \\
$\ \ \ \ \ C^{\dagger }C\left\vert \mathbf{\Lambda }_{n,m}\right\rangle $ & $%
=m\left\vert \mathbf{\Lambda }_{n,m}\right\rangle $ & , \\
$A^{\dagger }AC^{\dagger }C\left\vert \mathbf{\Lambda
}_{n,m}\right\rangle $
& $=nm\left\vert \mathbf{\Lambda }_{n,m}\right\rangle $ & .%
\end{tabular}
\label{B}
\end{equation}%
Taking the two Weyl spinors $\psi _{a}$ and $\bar{\xi}_{\dot{a}}$
like
\begin{equation}
\left(
\begin{array}{c}
\psi _{1} \\
\psi _{2} \\
\bar{\xi}_{1} \\
\bar{\xi}_{2}%
\end{array}%
\right) =\left(
\begin{array}{c}
\mathbf{\Lambda }_{n-1,\text{ }m}\text{ \ \ } \\
\mathbf{\Lambda }_{n,\text{ }m-1}\text{ \ } \\
\mathbf{\Lambda }_{n-1,\text{ }m-1} \\
\mathbf{\Lambda }_{n,\text{ }m}\text{ \ \ \ \ }%
\end{array}%
\right) \qquad \text{,}  \label{LA}
\end{equation}%
with remarkable chiral zero mode%
\begin{equation}
\left(
\begin{array}{c}
0\text{\ } \\
0 \\
0 \\
\mathbf{\Lambda }_{0,0}%
\end{array}%
\right)   \label{ZE}
\end{equation}%
and putting back into $\mathcal{OO}^{\dagger }\psi
_{a}=E_{n,m}^{2}\psi _{a}$
and $\mathcal{O}^{\dagger }\mathcal{O}\bar{\xi}_{\dot{a}}=E_{n,m}^{2}\bar{\xi%
}_{\dot{a}}$, we obtain the following the energy spectrum,%
\begin{equation}
E_{n,m}^{\pm }\left( \varpi ,\varpi ^{\prime }\right) =\pm \hbar \sqrt{%
n\varpi ^{2}+m\varpi ^{\prime 2}}.  \label{ENM}
\end{equation}%
This relation is comparable to (\ref{EN}) of graphene namely
$\varepsilon _{n}^{2D\text{-}graphene}=\pm \sqrt{\frac{2\hbar
e\mathcal{B}}{c}\left\vert
n\right\vert }$. Notice also that in the case of the background fields $%
\mathcal{B}\neq 0$; but $\mathcal{E}=0$, the tensor
$\mathcal{F}_{\mu \nu }$ is degenerate since $\det \mathcal{F}_{\mu
\nu }$ and the algebras of the
covariant derivatives (\ref{CR}) reduces to%
\begin{equation}
\left[ D_{1},D_{2}\right] =i\frac{Q\mathcal{B}}{c},\qquad D_{3}=\frac{%
\partial }{\partial z},\quad D_{4}=\frac{\partial }{\partial \tau },
\end{equation}%
and all remanning others which are just commutation relations
vanish.
Therefore one has only one quantum harmonic oscillators with frequency $%
\varpi $ whose energy spectrum is given by,
\begin{equation}
E_{n}^{\pm }\left( \varpi ,k\right) =\pm \hbar \sqrt{n\varpi
^{2}+k^{2}}
\end{equation}%
with $k^{2}=k_{3}^{2}+k_{4}^{2}$ and $p_{3}=\hbar k_{3},$
$p_{4}=\hbar k_{4}$ are the momenta along the z- and $\tau $-
directions. Notice that for the case $p_{3}=p_{4}=0$, the energy
levels are given by $E_{n}^{\pm }\left( \varpi ,k=0\right) =\pm
\hbar \varpi \sqrt{n}$ with $n$ a positive integer.
Notice also that for the fundamental state $n=0$, the energies vanish $%
E_{0}^{+}\left( \varpi ,k=0\right) =E_{0}^{-}\left( \varpi
,k=0\right) =0$;
and the valence and conducting bands touch; for an illustration see fig \ref%
{QH}.

\subsection{Filling factor $\protect\nu _{BC}$ of BC fermions in background
fields}

Before computing $\nu _{BC}$, we start by recalling that the wave function $%
\mathbf{\Psi }\left( \mathbf{x},t\right) $ of the \emph{BC} fermions
in the
background fields (\ref{FAC}) is given by%
\begin{equation}
\mathbf{\Psi }\left( \mathbf{x},t\right) =\sum_{n,m=-\infty }^{\infty }e^{-%
\frac{i}{\hbar }E_{n,m}t}\Psi _{n,m}\left( \mathbf{x}\right)
\label{EX}
\end{equation}%
with wave modes $\Psi _{n,m}=\left( \psi _{n,m}^{a},\bar{\xi}_{n,m}^{\dot{a}%
}\right) $ and energies $E_{n,m}$ respectively given by eqs(\ref{LA}) and (%
\ref{ENM}). Since the waves $\Psi _{n,m}$ form a basis and using
probability interpretation of $\mathbf{\Psi }\left(
\mathbf{x},t\right) $ we also have,
\begin{equation}
\int_{\mathbb{R}^{4}}d^{4}x\Psi _{n,m}^{\ast }\left(
\mathbf{x}\right) \Psi _{n^{\prime },m^{\prime }}\left(
\mathbf{x}\right) =\delta _{n,n^{\prime }}\delta _{m,m^{\prime }}.
\end{equation}%
These relations can be also derived by help of well known
orthogonality of the waves functions $\theta _{n}\left( x,y\right) $
and $\theta _{m}^{\prime }\left( z,\tau \right) $ of the quantum
harmonic oscillators. Now, using these relations, one can compute
the filling factor $\nu _{BC}$ of the AQHE
of the Bor\c{c}i-Creutz fermions in the constant background fields $\mathcal{%
B}$ and $\mathcal{E}$. This factor is given as usual by $\frac{\mathcal{N}%
_{F}}{\mathcal{N}_{\mathcal{\phi }}}$; that is the ratio of the number $%
\mathcal{N}_{F}$ of particle states with respect the number of
quantum fluxes $\mathcal{N}_{\phi }$. The latter number reads as
follows:
\begin{equation}
\mathcal{N}_{\mathcal{\phi }}=\frac{1}{2\pi }\int_{S^{2}}dx^{\mu
}\wedge dy^{\nu }\mathcal{F}_{\mu \nu }.
\end{equation}%
and, by using eq(\ref{FAC}), can be split as $\mathcal{N}_{\mathcal{B}%
}=\int_{S^{2}}\mathcal{B}dx\wedge dy$ and $\mathcal{N}_{\mathcal{E}%
}=\int_{S^{2}}\mathcal{E}dz\wedge d\tau $. For a unit flux $\mathcal{N}_{%
\mathcal{\phi }}=1$, the number $N_{F}$ of polarized particles that
fill the
band energy $0\leq E_{n,m}^{2}\leq E_{N,M}^{2}$ is given by the relation%
\begin{equation}
\mathcal{N}_{F}=\frac{1}{2}\int_{E_{N,M}^{-}\leq E_{n,m}\leq
E_{N,M}^{+}}dtd^{4}x\left\vert \boldsymbol{\Psi }\left(
\mathbf{x},t\right) \right\vert ^{2}
\end{equation}%
Substituting $\boldsymbol{\Psi }\left( \mathbf{x},t\right) $ by
eq(\ref{EX}) and using the orthogonality properties, we get
\begin{equation}
\mathcal{N}_{F}=\frac{\left( 2N+1\right) \left( 2M+1\right) }{2}.
\end{equation}%
In the particular case $N=M=0$, the energies $E_{0,0}^{-}$ and
$E_{0,0}^{+}$
vanish; and the conducting and valence coincide; so the fundamental state $%
\Psi _{0,0}$ with zero energy is filled by $\frac{1}{2}$ particle and $\frac{%
1}{2}$ hole. The number of polarized particles filling the band energy $%
0\leq E_{n,m}^{+}\leq E_{N,M}^{+}$ is equal to $\mathcal{N}_{F}$.
Therefore, taking into account the fact that a \emph{BC} fermion has
four spin polarizations; and each fermion has three colors, the
general expression of
the filling factor $\nu _{BC}$ reads as follow:%
\begin{equation}
\nu _{BC}=12k_{V}\frac{\left( 2N+1\right) \left( 2M+1\right) }{2},
\end{equation}%
where $k_{V}$ stands for the number of Dirac valleys which, for BC
fermions, is equal to $2$ as in \emph{2D} graphene.

\subsection{Link with spectral flow and topological index}

\begin{figure}[tbph]
\begin{center}
\hspace{0cm} \includegraphics[width=14cm]{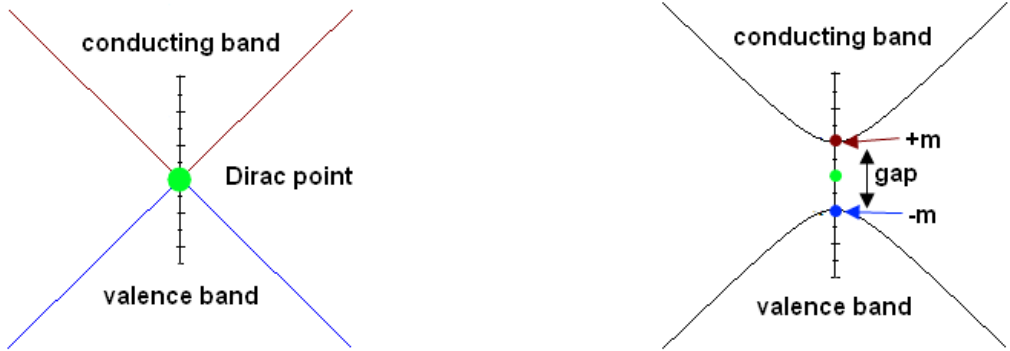}
\end{center}
\par
\vspace{-1cm} \caption{On left the Dirac point where conducting and
valence bands touch. On right the gap between {\protect\small
conducting and valence bands induced by the mass term.}} \label{GAP}
\end{figure}

In this subsection, we want to comment on the relation between the
hamiltonian, that we have used in this study and which has been
motivated from 2D graphene, and the so called spectral flow
hamiltonian considered recently in \textrm{\cite{4K}} in connection
with the theoretical foundation for the index theorem for fermions
on lattices. More analysis on this matter and other issues are
further developed in the appendices A and B.\newline In studying the
AQHE in 4-dimensions, we used the equation (\ref{DIR}) describing
the dynamics of a fermion near a generic Dirac point corresponding \
to a zero mode solving $D\Psi =0$. This equation, that we put in the
following simple form,
\begin{equation}
\left( D-E\right) \Psi =0,  \label{DE}
\end{equation}%
depends on the spectral parameter $E$, which in relativistic theory,
has a dimension of mass; it captures deformations away from the
Dirac point and allows to open the gap between the conducting and
the valence bands as shown on fig \ref{GAP}.
Recall that in eq(\ref{DE}), the Dirac operator in the background field $%
\mathcal{F}_{\mu \nu }$ as in eq(\ref{FAC}) is an antihermitian
operator
given by%
\begin{equation}
D=\gamma ^{\mu }D_{\mu }=\gamma ^{\mu }\left( \partial _{\mu
}-iA_{\mu }\right) .
\end{equation}%
Notice that the corresponding hamiltonian used in this study namely
$H\left( E\right) =\left( D-E\right) $ looks quite similar to the
one used in the method of spectral flow of \textrm{\cite{4K}}. There
the spectral flow
hamiltonian is given by \ the following%
\begin{equation}
H_{sp}\left( m\right) =\gamma _{5}\left( D-m\right)
\end{equation}%
with%
\begin{equation}
\gamma _{5}\left( D-m\right) \Psi =\lambda \left( m\right) \Psi ,
\end{equation}%
and the eigenvalues $\lambda \left( m\right) $ depending on the
spectral parameter m. Clearly the hamiltonian $H\left( E\right) $
and the spectral one $H_{sp}\left( m\right) $ are close cousins and
are related as follows,
\begin{equation}
H_{sp}\left( m\right) =\gamma _{5}H\left( m\right) .
\end{equation}%
The trick of multiplication by $\gamma _{5}$ has two \ nice and
remarkable effects; it makes $H_{sp}\left( m\right) $ hermitian and
allows to lift the degeneracy of energies with opposite chiralities.
Indeed, by splitting the Dirac spinor $\Psi $ into its two chiral
components $\Psi _{\pm }$ like
\begin{equation}
\begin{tabular}{lllll}
$\Psi _{+}=\left(
\begin{array}{c}
\phi  \\
0%
\end{array}%
\right) $ & , & $\Psi _{-}=\left(
\begin{array}{c}
0 \\
\chi
\end{array}%
\right) $ & , & $\gamma _{5}\Psi _{\pm }=\pm \Psi _{\pm }$%
\end{tabular}%
\end{equation}%
and restricting the study to the Dirac zero modes $D\Psi =0$, the
eigenvalue equation of the spectral flow hamiltonian reads as
follows
\begin{equation}
\begin{tabular}{ll}
$H_{sp}\left( m\right) \Psi _{\pm }=\mp m\Psi _{\pm }$ & ,%
\end{tabular}%
\end{equation}%
and captures perfectly the data of fig \ref{GAP} contrary to
$H\left(
E\right) $ which fails in this matter since we have%
\begin{equation}
\begin{tabular}{ll}
$H\left( m\right) \Psi _{\pm }=-m\Psi _{\pm }$ & .%
\end{tabular}%
\end{equation}%
It happens that the spectral flow hamiltonian $H_{sp}\left( m\right)
$ is a powerful ingredient as it leads to exactly the topological
index $Ind\left( D\right) $ of the Dirac operator. Following
\textrm{\cite{4K,2BB}} and the analysis developed in the appendices
A and B, one can show that the spectrum of $H_{sp}\left( m\right) $
is, up to an irrelevant sign, the topological
index%
\begin{equation}
Tr\left( \gamma _{5}e^{-tD}\right) =N_{+}-N_{-}.
\end{equation}%
In this relation, $N_{+}$ is the number of zero modes with positive
chirality; and $N_{-}$ is the number of zero modes with negative
chirality. The difference $N_{+}-N_{-}$ is equal to the topological
charge $Q_{top}$ of the gauge field; i.e: $N_{+}-N_{-}=Q_{top}$. For
the case of the hamiltonian $H\left( m\right) $, the corresponding
quantity (restricted to the zero modes) is given by
\begin{equation}
Tr\left( e^{-tD}\right) |_{\text{zero modes}}=N_{+}+N_{-},
\end{equation}%
it is not a topological quantity; it gets extra contribution from
\emph{non
zero} modes. We end by noting that in the present 4D case, we have%
\begin{equation}
\begin{tabular}{l|ll|l|ll}
quarks $q$ & $n_{q+}$ & $n_{q-}$ & degeneracy & $N_{q+}$ & $N_{q-}$
\\ \hline
&  &  &  &  &  \\
$u\left( x\right) $ & $0$ & $1$ & $\left\vert Q_{x}\right\vert
\left\vert Q_{z}\right\vert $ & $0$ & $\left\vert Q_{x}\right\vert
\left\vert
Q_{z}\right\vert $ \\
$d\left( x\right) $ & $0$ & $1$ & $\left\vert Q_{x}\right\vert
\left\vert Q_{z}\right\vert $ & $0$ & $\left\vert Q_{x}\right\vert
\left\vert
Q_{z}\right\vert $ \\
&  &  &  &  &  \\ \hline
\end{tabular}%
\end{equation}%
where the integer $\left\vert Q_{x}\right\vert \left\vert
Q_{z}\right\vert $ is the degeneracy of the zero modes; see
eq(\ref{36}-\ref{37}) of appendix A. The index theorem reads
therefore $N_{+}-N_{-}=$ $\left( N_{u+}+N_{d+}\right) -\left(
N_{u-}+N_{d-}\right) $ $=2\left\vert Q_{x}\right\vert \left\vert
Q_{z}\right\vert $ in agreement with results of \textrm{\cite{2BB}}.
More comments and technical details are reported in appendices A and
B.

\section{Conclusion and comments}

In this paper, we have studied the anomalous quantum Hall effect in
4D lattice QCD with a special focus on Bori\c{c}i-Creutz fermions in
presence of a constant background field strength $\mathcal{F}_{\mu
\nu }$. Recall that \emph{BC} fermions were proposed to simulate the
dynamics of light quarks using four dimensional lattice QCD and
where shown to be intimately linked the \emph{4D} extension of
graphene on hyperdiamond. We have shown that in the neighborhood of
the two Dirac valleys (\ref{ZM}), the \emph{BC} model has a
\emph{5}$\emph{D}$ field theoretical interpretation with the
following features:

\ \ \newline
(\textbf{1}) the background fields given by antisymmetric tensor $\mathcal{F}%
_{\mu \nu }$ appear as the $4\times 4$ submatrix of a \emph{5D}
field strength $\mathcal{F}_{MN}$ having $4+6$ components
distributed like
\begin{equation*}
\mathcal{F}_{MN}=\left(
\begin{array}{cc}
\mathcal{F}_{\mu \nu } & \mathcal{F}_{\mu 5} \\
-\mathcal{F}_{\mu 5} & 0%
\end{array}%
\right) .
\end{equation*}
In this \emph{5D} interpretation, the four components
$\mathcal{F}_{\mu 5}$ describing a 5D electric field are equal to
zero ($\mathcal{F}_{\mu 5}=0$); while the remaining \emph{6}
components $\mathcal{F}_{\mu \nu }$, which describe the components
of a \emph{5D} magnetic field, have been used to study the AQHE of
\emph{BC} fermions. This property explains also the behavior of the
background fields $\mathcal{B}$ and $\mathcal{E}$ as a magnetic
fields respectively rotating left and right fermions $\Psi_{L}$ and
$\Psi_{R}$; and leads to AQHE with cyclotron frequencies $\varpi =\sqrt{%
\frac{2Q\mathcal{B}}{c}}$ and $\varpi ^{\prime }=\sqrt{\frac{2Q\mathcal{E}}{c%
}}$.

\ \ \ \newline (\textbf{2}) the hamiltonian $H_{BC}$ is precisely
given by the euclidian
\emph{4D} Dirac operator $\sum_{\mu =1}^{4}i\gamma ^{\mu }D_{\mu }$ with $%
D_{\mu }=\left( \partial _{\mu }-i\frac{Q}{2c}\mathcal{F}_{\mu \nu
}x^{\nu }\right) $ capturing interactions between BC fermions and
the "\emph{5D} magnetic" tensor $\mathcal{F}_{\mu \nu }$. Because of
the non zero flux induced by the background field $\mathcal{F}_{\mu
\nu }$, the commuting algebra of the four flat space translations
$\partial _{\mu }$ turns into the non commutative $\left[ D_{\mu
},D_{\nu }\right] \sim i\mathcal{F}_{\mu \nu },$ $\left[ D_{\mu
},\mathcal{F}_{\nu \rho }\right] =0$. In the case where
$\mathcal{F}_{\mu \nu }$ is chosen as $\mathcal{B}\varepsilon _{\mu
\nu 34}+\mathcal{E}\varepsilon _{12\mu \nu }$, the algebra of the $D_{\mu }$%
's splits into two uncoupled Heisenberg ones,%
\begin{equation*}
\begin{tabular}{llll}
$\left[ D_{1},D_{2}\right] =-i\frac{Q\mathcal{B}}{2c}I$ & , &
$\left[
D_{3},D_{4}\right] =-i\frac{Q\mathcal{E}}{2c}I$ & ,%
\end{tabular}%
\end{equation*}%
whose spectrum as well as the associated AQHE have been explicitly
studied in this paper. This splitting teaches us moreover that
$\left[ D_{\mu },D_{\nu }\right] \sim i\mathcal{F}_{\mu \nu }$ and
describes in general two coupled quantum harmonic oscillators which
can be studied by using
perturbation theory if the following conditions hold%
\begin{equation*}
\begin{tabular}{ll}
$\left\vert \mathcal{F}_{14}+\mathcal{F}_{23}\right\vert
^{2}+\left\vert
\mathcal{F}_{13}-\mathcal{F}_{24}\right\vert ^{2}<<2\left\vert \mathcal{F}%
_{12}\mathcal{F}_{34}\right\vert $ & , \\
&  \\
$\left\vert \mathcal{F}_{14}-\mathcal{F}_{23}\right\vert
^{2}+\left\vert
\mathcal{F}_{13}+\mathcal{F}_{24}\right\vert ^{2}<<2\left\vert \mathcal{F}%
_{12}\mathcal{F}_{34}\right\vert $ & .%
\end{tabular}%
\end{equation*}%
We conclude this work by giving a comment on the extension of the
result given in this study to the case where the background gauge
potential is valued in the Cartan subalgebra of $SU_{c}\left(
3\right) \times
U_{em}\left( 1\right) $,%
\begin{equation*}
U^{2}\left( 1\right) \times U_{em}\left( 1\right) \subset
SU_{c}\left( 3\right) \times U_{em}\left( 1\right) .
\end{equation*}%
This subalgebra is a 3- dimensional abelian algebra generated, in
addition
to $h_{U_{em}\left( 1\right) }=Q_{em}I$, by the two diagonal hermitian $%
3\times 3$ matrices $h_{1}$ and $h_{2}$ of $SU_{c}\left( 3\right) $
which
read in the Gell-Mann vector basis as follows%
\begin{equation*}
\begin{tabular}{llllll}
$h_{em}=\left(
\begin{array}{ccc}
1 & 0 & 0 \\
0 & 1 & 0 \\
0 & 0 & 1%
\end{array}%
\right) $ &  & $h_{1}=\left(
\begin{array}{ccc}
1 & 0 & 0 \\
0 & -1 & 0 \\
0 & 0 & 0%
\end{array}%
\right) $ & , & $h_{2}=\frac{1}{\sqrt{3}}\left(
\begin{array}{ccc}
1 & 0 & 0 \\
0 & 1 & 0 \\
0 & 0 & -2%
\end{array}%
\right) $ & .%
\end{tabular}%
\end{equation*}%
In presence of non zero constant background fields $\mathcal{F}_{\mu
\nu }^{\left( em\right) }\neq 0,$ $\mathcal{F}_{\mu \nu }^{\left(
1\right) }\neq 0,$ $\mathcal{F}_{\mu \nu }^{\left( 2\right) }\neq
0$, the gauge covariant
derivatives reads as follows,%
\begin{equation*}
D_{\mu }=\partial _{\mu }-x^{\nu }\left(
\frac{ig}{2c}h_{1}\mathcal{F}_{\nu \mu }^{\left( 1\right)
}+\frac{ig}{2c}h_{2}\mathcal{F}_{\nu \mu }^{\left( 2\right)
}+\frac{iQ_{em}}{2c}\mathcal{F}_{\nu \mu }^{\left( em\right)
}\right)
\end{equation*}%
obeying the non commutative commutation relations $\left[ D_{\mu },D_{\nu }%
\right] \sim i\mathcal{F}_{\mu \nu }$; but now with
$\mathcal{F}_{\mu \nu
}=\sum_{I}h_{I}\mathcal{F}_{\mu \nu }^{I}$ that reads in matrix notation like%
\begin{equation*}
\left(
\begin{array}{ccc}
\mathcal{F}_{\mu \nu }^{\left( em\right) }+\mathcal{F}_{\mu \nu
}^{\left(
1\right) }+\mathcal{F}_{\mu \nu }^{\left( 2\right) } & 0 & 0 \\
0 & \mathcal{F}_{\mu \nu }^{\left( em\right) }-\mathcal{F}_{\mu \nu
}^{\left( 1\right) }+\mathcal{F}_{\mu \nu }^{\left( 2\right) } & 0 \\
0 & 0 & \mathcal{F}_{\mu \nu }^{\left( em\right) }-2\mathcal{F}_{\mu
\nu
}^{\left( 2\right) }%
\end{array}%
\right)
\end{equation*}%
Moreover, because the $h_{I}$'s are diagonal matrices; each gauge
covariant derivative $D_{\mu }$ splits into $\emph{3}$ components
$D_{\mu }^{\left( 1\right) }$, $D_{\mu }^{\left( 2\right) }$,
$D_{\mu }^{\left( 3\right) }$ as given below,
\begin{equation*}
\begin{tabular}{lll}
$D_{\mu }^{\left( 1\right) }$ & $=\partial _{\mu
}-\frac{i}{2c}x^{\nu }\left( Q_{em}\mathcal{F}_{\nu \mu }^{\left(
em\right) }+g\mathcal{F}_{\mu \nu }^{\left( 1\right)
}+g\mathcal{F}_{\mu \nu }^{\left( 2\right) }\right) $
&  \\
$D_{\mu }^{\left( 2\right) }$ & $=\partial _{\mu
}-\frac{i}{2c}x^{\nu }\left( Q_{em}\mathcal{F}_{\nu \mu }^{\left(
em\right) }-g\mathcal{F}_{\mu \nu }^{\left( 1\right)
}+g\mathcal{F}_{\mu \nu }^{\left( 2\right) }\right) $
&  \\
$D_{\mu }^{\left( 3\right) }$ & $=\partial _{\mu
}-\frac{i}{2c}x^{\nu }\left( Q_{em}\mathcal{F}_{\nu \mu }^{\left(
em\right) }-2g\mathcal{F}_{\mu \nu }^{\left( 2\right) }\right) $ &
\end{tabular}%
\end{equation*}%
and one is left with \emph{6} quantum harmonic oscillators whose
uncoupled realization is given by the choice
\begin{equation*}
\mathcal{F}_{\mu \nu }^{I}=\mathcal{B}^{I}\varepsilon _{\mu \nu 34}+\mathcal{%
E}^{I}\varepsilon _{12\mu \nu }.
\end{equation*}%
In this case the filling factor of the AQHE of BC fermions induced by these $%
\mathcal{F}_{\mu \nu }^{I}$'s is given by eq(\ref{FL}). In the end,
it would be interesting to study the analogue of the Zeeman effect
of \emph{2D} graphene in 4D lattice QCD by using models like the
\emph{BC} fermions we have considered in this study.\newline We
close this conclusion by two more comments. The first comment
concerns the relation between the our results and the index theorem.
This matter has been described in a condensed form throughout the
paper; in particular in subsection 5.3. It is developed with details
in appendices A and B. The second comment concerns QHE in lattice
QCD as whole; the present study focused in BC fermions should be
viewed as a first step which itself need further investigations;
several issues like the explicit breaking of discrete symmetries in
BC fermions \textrm{\cite{2BC}} as well as spin Hall effect and the
associated topological indices \textrm{\cite{Z,ZZ}} haven't been
addressed here.

\section{Appendix A: Index of Dirac operator}

In this section, we describe the index theorem of the Dirac operator
\emph{D} in background gauge configuration. This index provides the
relationship between physical properties and the topology of the
space in which live the fermions. First, we describe the index in
2-dimensions; both in continuum and lattice QFTs. Then, we consider
the main lines of the index in 4-dimensions and its relationship
with the so called spectral flow hamiltonian. After that, we make an
heuristic interpretation of the filling factors obtained in this
study in terms of the index of the Dirac operator.

\subsection{case of 2D graphene}

We begin by describing the topological index theorem of the Dirac
operator in 2-dimensions. Then we apply the construction to the case
of \emph{2D} graphene.

\subsubsection{the index theorem: continuous limit}

In the case of fermions living on a 2-dimensional surface in
presence of external gauge fields, the index theorem gives a
relation between zero modes of the gauged Dirac operator
$\mathcal{D}=\sigma _{\alpha \beta }^{\mu }D_{\mu }$, with gauge
covariant derivative $D_{\mu }=\left( \partial _{\mu }-iA_{\mu
}\right) $, and the flux $Q=\frac{1}{2\pi }\int_{S}\mathcal{F}_{2}$
going through $\mathcal{S}$
($\mathcal{F}_{2}=\frac{1}{2}\mathcal{F}_{\mu \nu }dx^{\mu }\wedge
dx^{\nu }$). Writing this Dirac operator as $2\times 2$
matrix $\mathcal{D}$ as follows,%
\begin{equation}
\mathcal{D}=\left(
\begin{array}{cc}
0 & D_{+} \\
D_{-} & 0%
\end{array}%
\right) ,\qquad \left.
\begin{array}{c}
D_{+}=D_{1}+iD_{2} \\
D_{-}=D_{1}-iD_{2}%
\end{array}%
\right. ,\qquad \left( D_{+}\right) ^{\dagger }=-D_{-}
\end{equation}%
and the flux $Q$ through $\mathcal{S}$ like $Q=\frac{BS}{2\pi }$ with $B=%
\frac{1}{2}\varepsilon ^{3\mu \nu }\mathcal{F}_{\mu \nu }$, the
index of the
Dirac operator reads explicitly in terms of the number $N_{0}^{+}$ ($%
N_{0}^{-}$) of zero modes with positive (negative) chirality and the
magnetic flux as
\begin{equation}
N_{0}^{+}-N_{0}^{-}=Q.  \label{q}
\end{equation}%
This index is defined like $Ind\left( \mathcal{D}\right) =Tr\left(
\sigma ^{3}e^{-t\lambda \mathcal{D}}\right) $ with $t$ some spectral
parameter and leads to the above equality which can be got by
computing $Ind\left( \mathcal{D}\right) $ in two different ways and
equate the two results. Let us describe rapidly how this is done.
\newline
(1) use the property that the Dirac operator $\mathcal{D}$ and its square $%
\mathcal{D}^{2}$ have the same number of zero modes. Since
$\mathcal{D}^{2}$
has a diagonal form%
\begin{equation}
\mathcal{D}^{2}=\eta ^{\mu \nu }\mathcal{D}_{\mu }\mathcal{D}_{\nu }-\frac{i%
}{4}\left[ \sigma ^{\mu },\sigma ^{\nu }\right] \mathcal{F}_{\mu \nu
}
\end{equation}%
as clearly seen on the matrix representation
\begin{equation}
\begin{tabular}{llll}
$\mathcal{D}^{2}=\left(
\begin{array}{cc}
D_{+}D_{-} & 0 \\
0 & D_{-}D_{+}%
\end{array}%
\right) $ & , & $\mathcal{D}^{2}\Psi =E^{2}\Psi $ & ,%
\end{tabular}
\label{D2}
\end{equation}%
one can do explicit calculations by using this property. Let show
rapidly how this works. From above equation, we learn that the
operators $D_{+}D_{-}$ and $D_{-}D_{+}$ have the same \emph{non
zero} eigenvalues $E\neq 0$; this
means that they have the number of \emph{non zero} modes $N_{n}^{+}$ and $%
N_{n}^{-}$
\begin{equation}
N_{n}^{+}=N_{n}^{-},\qquad n\neq 0,
\end{equation}%
with $\pm $ refereing to positive and negative chiralities. Denoting by $%
\Psi =\left( \Psi _{+},\Psi _{-}\right) $ the 2D spinor with $\Psi
_{+}$ and $\Psi _{-}$ standing for the positive and negative
chiralities; and assuming that
\begin{equation}
\begin{tabular}{lll}
$D_{+}D_{-}\Psi _{+}=\lambda \Psi _{+}$ & , & $\lambda \neq 0$%
\end{tabular}
\label{D}
\end{equation}%
then $D_{-}\Psi _{+}$ is an eigenstate of $D_{-}D_{+}$ with the same
eigenvalue; but with opposite chirality. This feature follows
directly by
multiplying both sides of eq(\ref{D}) by $D_{-}$. However in the case where $%
\lambda =0$; that is $D_{+}D_{-}\Psi _{+}=0$, the equality $%
N_{n}^{+}=N_{n}^{-}$ doesn't necessary hold for zero modes since
$D_{-}\Psi _{+}$ may be itself equal to zero; so
$N_{0}^{+}-N_{0}^{-}$ may be different
from zero. Moreover, using (\ref{D2}), it is not difficult to see that%
\begin{equation}
\begin{tabular}{llll}
$Tr\left( \tau ^{3}e^{-t\mathcal{D}^{2}}\right) $ & $=$ & $Tr\left(
e^{-tD_{+}D_{-}}\right) -Tr\left( e^{-tD_{-}D_{+}}\right) $ & .%
\end{tabular}%
\end{equation}%
Expanding the trace in terms of zero modes and non zero modes the
above relation can be also put in the form
\begin{equation*}
\begin{tabular}{ll}
$Ind\left( \mathcal{D}\right) =\left( \dsum\limits_{E_{+}^{\left(
=0\right) }}e^{-tE_{+}^{\left( =0\right)
}}-\dsum\limits_{E_{-}^{\left( =0\right) }}e^{-tE_{-}^{\left(
=0\right) }}\right) +\left( \dsum\limits_{E_{+}^{\left( \neq
0\right) }}e^{-tE_{+}^{\left( \neq 0\right)
}}-\dsum\limits_{E_{-}^{\left( \neq 0\right) }}e^{-tE_{-}^{\left(
\neq
0\right) }}\right) $ & ,%
\end{tabular}%
\end{equation*}%
leading afterwards to%
\begin{equation}
Tr\left( \tau ^{3}e^{-t\mathcal{D}^{2}}\right) =N_{0}^{+}-N_{0}^{-}.
\label{N0}
\end{equation}%
(2) To get the right hand side of eq(\ref{q}), we use two more
features of
the index; first the remarkable independence of $Tr\left( \tau ^{3}e^{-t%
\mathcal{D}^{2}}\right) $ on the spectral parameter $t$ as
explicitly exhibited by (\ref{N0}). Second use the so called
\emph{heat expansion}
method \textrm{\cite{P} }%
\begin{equation}
Ind\left( \mathcal{D}\right) =\frac{1}{4\pi t}\dsum\limits_{n\geq 0}t^{\frac{%
n}{2}}b_{n}\left( \mathcal{D}\right)
\end{equation}%
where the $b_{n}\left( \mathcal{D}\right) $'s are expansion
coefficients. Since this expression should be t-invariant; it
follows that $Ind\left( \mathcal{D}\right) =\frac{b_{2}\left(
\mathcal{D}\right) }{4\pi }$; which by
using eq(\ref{K2}) leads to%
\begin{equation}
b_{2}\left( \mathcal{D}\right) =Tr\left[ \sigma ^{3}\left(
\frac{-i}{4}\left[ \sigma ^{\mu },\sigma ^{\nu }\right]
\mathcal{F}_{\mu \nu }\right) \right] =2\int BdS
\end{equation}%
with $\left[ \sigma ^{\mu },\sigma ^{\nu }\right] =2i\varepsilon
^{\mu \nu 3}\sigma ^{3}$ and $\frac{1}{2\pi }\int BdS=Q$ producing
the total magnetic monopole charge in discrete values inside the
surface.

\subsubsection{2D fermions on lattice}

To get the topological index for 2D graphene, one starts from the
analysis given in previous subsection; then works out the extension
fermions on
honeycomb. The index on a square lattice has been studied in \textrm{\cite{S}%
}; and the one on honeycomb has been considered recently in \textrm{\cite{SR}%
}. Below we give a brief description of the main lines of the two
constructions.

\emph{fermions on a square lattice}\newline
In the case of a finite $L\times L^{\prime }$ square lattice where $L=Na$, $%
L^{\prime }=N^{\prime }a$ with $a$ the spacing lattice parameter and
individual $\mathcal{S}_{square}=a^{2}$, the gauge configuration is
chosen
as follows:%
\begin{equation}
\begin{tabular}{llll}
$A_{1}\left( x,y\right) =-\omega y$, & $A_{2}\left( x,y\right) =0,$ & $%
F_{12} $ & $=B$%
\end{tabular}%
\end{equation}%
with flux%
\begin{equation}
\begin{tabular}{lll}
$Q$ & $=\frac{1}{2\pi }\int dxdyF_{12}=\frac{BS}{2\pi }$ & .%
\end{tabular}%
\end{equation}%
which, for later use, we write it as follows,%
\begin{equation}
\begin{tabular}{lll}
$Q=\frac{BLL^{\prime }}{2\pi }$, & $B=\frac{2\pi }{LL^{\prime }}Q,$ & $%
\mathcal{S}=LL^{\prime }$,%
\end{tabular}%
\end{equation}%
We also have the following boundary conditions
\begin{equation}
\begin{tabular}{lllll}
$A_{1}|_{y=0}$ & \multicolumn{3}{l}{$=A_{1}|_{y=L^{\prime }}+i\Omega \frac{%
\partial }{\partial x}\Omega ^{-1}$} & $,$ \\
&  &  &  &  \\
$\Omega \left( x\right) $ & $=e^{iBL^{\prime }x}$ & , & $\Omega
\left(
x\right) |_{x=0}=\Omega \left( x\right) |_{y=L}$ & ,%
\end{tabular}%
\end{equation}%
leading to $BLL^{\prime }=2\pi n$ with n integer. Notice that the
discontinuity in the vector potential is captured by a gauge
transformation.
Notice also that the periodicity condition on the transition function $%
\Omega \left( x\right) $ requires $\frac{B}{2\pi }\mathcal{S}=Q\in
\mathbb{Z} $; it teaches us that the field strength $B$ and the
topological charge $Q$ are quantized. By following
\textrm{\cite{S}}, the general solution of the Dirac equation of the
2-dimensional fermions on the square lattice
satisfying the above boundary conditions is given by%
\begin{equation}
\begin{tabular}{llll}
$\Psi _{n}^{\pm }\left( x,y\right) =\dsum\limits_{k\in Z}\mathcal{C}%
_{k}^{\pm }e^{i\frac{2\pi k}{L}x}e^{-\frac{1}{2}\xi ^{2}}H_{n}\left(
\xi
\right) $ & , & $\xi =\sqrt{\left\vert B\right\vert }\left( y+\frac{k}{Q}%
L^{\prime }\right) $ &
\end{tabular}%
\end{equation}%
where $H_{n}\left( \xi \right) $ are Hermite polynomials of order
$n$; and where the $\mathcal{C}_{k}^{\pm }$ coefficients are
constrained by the recurrent relations $\mathcal{C}_{k}^{\pm
}=\mathcal{C}_{k-Q}^{\pm }$; which result from boundary conditions
and showing that only $\mathcal{C}_{0}^{\pm
},...,\mathcal{C}_{\left\vert Q\right\vert -1}^{\pm }$ which can be
chosen arbitrary.\ The eigenvalues associated with the $\Psi
_{n}^{\pm }\left(
x,y\right) $'s are given by%
\begin{equation}
\begin{tabular}{lll}
$\left( E_{n}^{+}\right) ^{2}=$ & $2\left( n+1\right) \left\vert
B\right\vert -B$ & , \\
$\left( E_{n}^{-}\right) ^{2}=$ & $2\left( n+1\right) \left\vert
B\right\vert +B$ & .%
\end{tabular}%
\end{equation}%
The link between chirality and zero modes depends on the sign of the
topological charge $Q$ or the sense of the external magnetic field
$B$. For the case $B>0$, the above relations read as $\left(
E_{n}^{+}\right) ^{2}=2nB $ and $\left( E_{n}^{+}\right)
^{2}=2\left( n+1\right) B$. So only the wave function with positive
chirality has zero modes with degree of degeneracy equal to
$\left\vert Q\right\vert $. For the case $B<0$, the zero
modes have a negative chirality with $\left\vert Q\right\vert $ degeneracy.%
\newline
For the case of $\mathcal{N}$ continuum flavors described by the
Dirac operator in the magnetic background field as specified above,
the index
theorem reads therefore as follows%
\begin{equation}
Ind\left( D_{square}\right) =\mathcal{N}\left\vert Q\right\vert .
\end{equation}%
As such, the index of the Dirac operators for minimally doubled
fermions; in particular for BC fermions is $2\left\vert Q\right\vert
$. For naive fermions it is equal to $16$ $\left\vert Q\right\vert
$.

\emph{honeycomb fermions} \newline On the honeycomb, the situation
is quite similar; the main difference comes form the
crystallographic structure. The honeycomb is given by the
superposition of two sublattices $\mathcal{A}_{gra}$ and $\mathcal{B}_{gra}$%
. An interesting way to parameterize these sublattices is in terms
of the
two simple roots $\mathbf{\alpha }_{1}$ and $\mathbf{\alpha }_{2}$ of $%
SU\left( 3\right) $ and the weight vectors $\mathbf{\lambda }_{1},$ $\mathbf{%
\lambda }_{2},$ $\mathbf{\lambda }_{3}$ of its fundamental
representations
\textrm{\cite{SR,2C,DR,RD}}. More precisely sites $\mathbf{r}_{n}$ in $%
\mathcal{A}$ and $\mathbf{r}_{n}^{\prime }$ in $\mathcal{B}$ are
expanded as
follows%
\begin{equation}
\begin{tabular}{llll}
$\mathcal{A}:$ & $\frac{\mathbf{r}_{n}}{d}=n_{1}\mathbf{\alpha }_{1}+n_{2}%
\mathbf{\alpha }_{2}$ & $,$ & $n=\left( n_{1}\mathbf{,}n_{2}\right)
\in
\mathbb{Z}^{2}$ \\
$\mathcal{B}:$ & $\mathbf{r}_{n}^{\prime }=\mathbf{r}_{n}+\mathbf{s}$ & , & $%
d=a\frac{\sqrt{3}}{2}$%
\end{tabular}%
\end{equation}%
where $\mathbf{\alpha }_{1}$, $\mathbf{\alpha }_{2}$ and their sum $\mathbf{%
\alpha }_{3}=\mathbf{\alpha }_{1}+\mathbf{\alpha }_{2}$ which is
also a root
but not simple; can be taken like%
\begin{equation}
\begin{tabular}{llllll}
$\mathbf{\alpha }_{1}=(\frac{\sqrt{6}}{2},-\frac{\sqrt{2}}{2})$ & , & $%
\mathbf{\alpha }_{2}=(0,\sqrt{2})$ & , & $\mathbf{\alpha }_{3}=(\frac{\sqrt{6%
}}{2},\frac{\sqrt{2}}{2})$ & ,%
\end{tabular}%
\end{equation}%
satisfying the usual root's properties of $SU\left( 3\right) $; in
particular $\mathbf{\alpha }_{i}^{2}=2$. Moreover, the shift vector $\mathbf{%
s}$ is related to one of the weight vectors like $\mathbf{\lambda }=\mathbf{s%
}\frac{\sqrt{6}}{3}$; it can be any one of the three following
\begin{equation}
\begin{tabular}{llllll}
$\mathbf{s}_{1}=a(1,0)$ & , & $\mathbf{s}_{2}=a(-\frac{1}{2},\frac{\sqrt{3}}{%
2})$ & , & $\mathbf{s}_{3}=a(-\frac{1}{2},-\frac{\sqrt{3}}{2})$ & .%
\end{tabular}%
\end{equation}%
Notice that the sum
$\mathbf{s}_{1}+\mathbf{s}_{2}+\mathbf{s}_{3}=0$, its
captures the traceless property of the fundamental representation of SU$%
\left( 3\right) $; we also have the relations
\begin{equation}
\begin{tabular}{llll}
$\left( \mathbf{s}_{1}-\mathbf{s}_{2}\right) =\sqrt{\frac{3}{2}}\mathbf{%
\alpha }_{1},$ & $\left( \mathbf{s}_{2}-\mathbf{s}_{3}\right) =\sqrt{\frac{3%
}{2}}\mathbf{\alpha }_{2},$ & $\left( \mathbf{s}_{3}-\mathbf{s}_{1}\right) =%
\sqrt{\frac{3}{2}}\mathbf{\alpha }_{0}$ & .%
\end{tabular}%
\end{equation}%
with $\mathbf{\alpha }_{0}=-\mathbf{\alpha }_{3}$. Plaquettes in the \emph{2D%
} honeycomb are hexagonal with area $\mathcal{S}_{hexa}=a^{2}\frac{\sqrt{3}}{%
2}$; so the magnetic field is quantized as follows%
\begin{equation}
B=\frac{4\pi }{NN^{\prime }a^{2}\sqrt{3}}Q.
\end{equation}%
In \textrm{\cite{SR}}, the two lattice axes of the honeycomb were
chosen as
generated by the roots $\left( \mathbf{\alpha }_{0},-\mathbf{\alpha }%
_{1}\right) $, i.e: $\mathbf{r}=x_{0}\mathbf{\alpha }_{0}-x_{1}\mathbf{%
\alpha }_{0}$; and the boundary conditions to have a finite
translationally
invariant 2D graphene lattice $L_{0}\times L_{1}$ were taken like%
\begin{equation}
\begin{tabular}{lll}
$F\left( \mathbf{r}+L_{0}\mathbf{\alpha }_{0}\right) $ & $=F\left( \mathbf{r}%
\right) $ & , \\
$F\left( \mathbf{r}+L_{1}\mathbf{\alpha }_{1}\right) $ & $=F\left( \mathbf{r}%
\right) $ & .%
\end{tabular}%
\end{equation}%
Moreover, the link field configuration depending on these boundary
conditions are given by%
\begin{equation}
\begin{tabular}{llll}
\multicolumn{3}{l}{$U\left( \mathbf{r,s}_{1}\right) =e^{-i\frac{Ba^{2}\sqrt{3%
}}{2}x_{0}}$} & , \\
$U\left( \mathbf{r,s}_{2}\right) =1$ & $,$ & $U\left( \mathbf{r,s}%
_{2}\right) =1$ & ,%
\end{tabular}%
\end{equation}%
for all cells except those of the last row with $x_{0}=L_{0}-1$
where it is
required moreover%
\begin{equation}
\begin{tabular}{ll}
$U\left( x_{0}=L_{0}-1,x_{1}\mathbf{,s}_{3}\right) =e^{-i\frac{Ba^{2}\sqrt{3}%
}{2}L_{0}x_{1}}$ & .%
\end{tabular}%
\end{equation}%
Like in the square lattice, the topological index is also given by $%
Ind\left( D_{gra}\right) =2\left\vert Q\right\vert $; the main
difference is that it is given by $\Psi ^{+}\Sigma ^{3}\Psi $ where
$\Sigma ^{1}$, $\Sigma ^{2}$, $\Sigma ^{3}$ are the generators of
the $SU\left( 2\right) $ flavor
symmetry rotating the two Dirac points; for an explicit analysis see \textrm{%
\cite{SR}.}

\subsection{Index in 4D\ lattice QCD}

In 4-dimensions, the determination of the index of the Dirac
operator of fermions living on \emph{4D} spaces follows the same
approach as in \emph{2D} case. Let us describe briefly the main
lines of the method.

\subsubsection{Fermions on 4D space}

To get the topological index in a $SU\left( N\right) $ background
gauge
field configuration one has to compute both sides of the 4D analogue of eq(%
\ref{q}) namely
\begin{equation}
N_{0}^{+}-N_{0}^{-}=2\left\vert Q\right\vert ,  \label{NQ}
\end{equation}%
where $N_{0}^{+}$ are the number of chiral zero modes and $Q$ the
flux. Let us compute the topological charge $Q$ that gives the right
hand side of this relation. It is generally given by:
\begin{equation}
Q=\frac{1}{32\pi ^{2}}\int_{S_{4}}d^{4}x\varepsilon _{\mu \nu \rho
\sigma }Tr\left( \mathcal{F}^{\mu \nu }\mathcal{F}^{\rho \sigma
}\right) , \label{FFF}
\end{equation}%
with field strength $\mathcal{F}_{\mu \nu }=\partial _{\mu }A_{\nu
}-\partial _{\nu }A_{\mu }-i\left[ A_{\mu },A_{\nu }\right] $ valued
in the adjoint representation of $SU\left( N\right) $; i.e
$\mathcal{F}_{\mu \nu }=\sum_{a}T_{a}\mathcal{F}_{\mu \nu }^{a}$.
Making a simple choice of this gauge invariant field as follows
\begin{equation}
\mathcal{F}_{\mu \nu }=\left(
\begin{array}{cccc}
0 & B_{1} & 0 & 0 \\
-B_{1} & 0 & 0 & 0 \\
0 & 0 & 0 & B_{2} \\
0 & 0 & -B_{2} & 0%
\end{array}%
\right) \otimes T,  \label{F}
\end{equation}%
where T is one of the generators of $SU\left( N\right) $ normalized to Tr$%
\left( T^{2}\right) =2$, one can determine, up to a gauge
transformation, the corresponding gauge potentials. Notice that a
gauge configuration that
is appropriate to lattice computations is given by,%
\begin{equation}
\begin{tabular}{llll}
$A_{1}\left( x,y,z,\tau \right) $ & $=-B_{1}yT$ & , & $A_{2}\left(
x,y,z,\tau \right) =0$ \\
$A_{3}\left( x,y,z,\tau \right) $ & $=-B_{2}zT$ & $,$ & $A_{4}\left(
x,y,z,\tau \right) =0$%
\end{tabular}%
\end{equation}%
Putting (\ref{F}) back into (\ref{FFF}) by using,%
\begin{equation}
\mathcal{\tilde{F}}^{\mu \nu }=\left(
\begin{array}{cccc}
0 & -B_{2} & 0 & 0 \\
B_{2} & 0 & 0 & 0 \\
0 & 0 & 0 & -B_{1} \\
0 & 0 & B_{1} & 0%
\end{array}%
\right) \otimes T,
\end{equation}%
which is (anti) self dual for $B_{2}=\pm B_{1}$, one gets the
explicit
expression of the topological charge $Q$ in terms of the fields $B_{1}$ and $%
B_{2}$ and the volume of the 4-dimensional compact hyper surface. We have,%
\begin{equation}
Q=\frac{16B_{1}B_{2}}{32\pi ^{2}}Vol\left( S_{4}\right) \in \mathbb{Z}\text{.%
}
\end{equation}

\subsubsection{4D hypercube}

This relation can be given a more explicit form by working by
considering fermions on lattices. In the case of the \emph{4D}
hypercube $L_{1}\times L_{2}\times L_{3}\times L_{4}$, the previous
relations reads as
\begin{equation}
\begin{tabular}{ll}
$Q=\frac{16B_{1}B_{2}}{32\pi ^{2}}\left(
\dprod\limits_{i=1}^{4}L_{i}\right)
$ & .%
\end{tabular}%
\end{equation}%
Moreover use the fact that the field strengths $B_{1}$ and $B_{2}$
are
quantized as%
\begin{equation}
\begin{tabular}{lll}
$B_{1}=\frac{2\pi N_{1}}{L_{1}L_{2}},$ & $B_{2}=\frac{2\pi N_{2}}{L_{3}L_{4}}%
,$ & $N_{1,2}\in \mathbb{Z}$,%
\end{tabular}%
\end{equation}%
we end with the following topological charge
\begin{equation}
Q=2N_{1}N_{2},
\end{equation}%
which is independent from the nature of the 4D lattice; and is then
also valid for the 4D hyperdiamond of \emph{4D} lattice QCD.
Regarding the left hand side of (\ref{NQ}); it is determined by
solving the Dirac equation. We
find%
\begin{equation}
\Psi _{n,m}^{\pm }\left( x,y\right) \dsum\limits_{\left(
k_{x},k_{z}\right) \in \mathbb{Z}^{2}}\mathcal{C}_{k_{x},k_{z}}^{\pm
}\exp i\left( \frac{2\pi k_{x}}{L_{1}}x+\frac{2\pi
k_{z}}{L_{3}}z\right) e^{-\frac{1}{2}\left( \xi ^{2}+\zeta
^{2}\right) }H_{n,m},
\end{equation}%
with%
\begin{equation}
\begin{tabular}{lll}
$H_{n,m}$ & $=H_{n}\left( \xi \right) \times H_{m}\left( \zeta
\right) $ &
\\
$\ \ \ \xi $ & $=\sqrt{\left\vert B_{1}\right\vert }\left( y+\frac{k_{x}}{%
Q_{x}}L_{2}\right) $ &  \\
$\ \ \ \zeta $ & $=\sqrt{\left\vert B_{2}\right\vert }\left( \tau +\frac{%
k_{z}}{Q_{z}}L_{4}\right) $ &
\end{tabular}
\label{36}
\end{equation}%
where $H_{n}\left( \xi \right) $ are Hermite polynomials; and where
the coefficients $\mathcal{C}_{k_{x},k_{z}}^{\pm }$ are constrained
by the recurrent relations
\begin{equation}
\mathcal{C}_{k_{x},k_{z}}^{\pm
}=\mathcal{C}_{k_{x}-Q_{x},k_{z}-Q_{z}}^{\pm },  \label{37}
\end{equation}%
showing that the degeneracy of the chiral zero mode is $\left\vert
Q_{x}Q_{z}\right\vert $.

\section{Appendix B: Index, spectral flow and filling factor}

\subsection{Index theorem and spectral flow}

In this section, we describe the main lines of the spectral flow
method to get the topological index which can be symbolically stated
as
\begin{equation}
Ind\left( D\right) =-\mathrm{spf}\left( \mathcal{H}\right)
\end{equation}%
where $\mathrm{spf}\left( H\right) $ stands for spectral flow of
$\mathcal{H}
$ which, reads in terms of the 4-dimensional Dirac operator $\mathcal{D}%
=\gamma ^{\mu }\left( \partial _{\mu }-iA_{\mu }\right) $, as follows%
\begin{equation}
\begin{tabular}{llll}
$\mathcal{H}\left( m\right) =\gamma _{5}\left( \mathcal{D}-m\right) $ & , & $%
m\in \mathbb{R}$ & .%
\end{tabular}%
\end{equation}%
This approach was first considered in\textrm{\ \cite{4K} }for
staggered lattice fermions where\textrm{\ }the would-be chiral
zero-modes has been
identified away from the continuum limit. Then, it has extended in \textrm{%
\cite{2BA,2BB}} to minimally doubled fermions as well as the naive
ones by using the point splitting method for implementing flavored
mass terms. The spectral flow method detects exactly the index of
the would-be zero modes fixing the gauge field topology; it has been
explicitly checked numerically for 2D and 4D staggered, minimally
doubled and naive fermions. Below, we will mainly focus on 4D; but
the results can be extended to all even dimensions. \newline The key
idea of the spectral flow method relies on defining a hermitian
spectral hamiltonian depending on two basic things. (1) The Dirac
equation
for zero modes namely%
\begin{equation}
\left(
\begin{array}{cc}
0 & D^{\dagger } \\
-D & 0%
\end{array}%
\right) \left(
\begin{array}{c}
\phi \\
\chi%
\end{array}%
\right) =0,
\end{equation}%
where each block is a $2\times 2$ matrix and where the operators $D$ and $%
D^{\dagger }$ are as in eq(\ref{RC},\ref{AAC}) and
\begin{equation}
\Psi =\left(
\begin{array}{c}
\phi \\
\chi%
\end{array}%
\right)
\end{equation}%
being the Dirac spinor in 4-dimensions with the two chiral
components $\Psi
_{\pm }$ given below,%
\begin{equation}
\begin{tabular}{lll}
$\Psi _{+}=\left(
\begin{array}{c}
\phi \\
0%
\end{array}%
\right) ,$ & $\Psi _{-}=\left(
\begin{array}{c}
0 \\
\chi%
\end{array}%
\right) ,$ & $\gamma _{5}\Psi _{\pm }=\pm \Psi _{\pm }.$%
\end{tabular}%
\end{equation}%
(2) the spectral hamiltonian operator $\mathcal{H}\left( m\right)
=\gamma _{5}\left( \mathcal{D}-m\right) $ depending on a real
spectral parameter $m$ which can be thought of as mass. In matrix
notation, we have,
\begin{equation}
\begin{tabular}{llllll}
$\mathcal{D}$ & $=\left(
\begin{array}{cc}
0 & D^{\dagger } \\
-D & 0%
\end{array}%
\right) $ & , & $\mathcal{H}$ & $=\left(
\begin{array}{cc}
m & D^{\dagger } \\
-D^{\dagger } & -m%
\end{array}%
\right) $ & ,%
\end{tabular}%
\end{equation}%
Notice also the traceless property of the spectral hamiltonian operator%
\begin{equation}
Tr\mathcal{H}=0,
\end{equation}%
which, a priori, should be independent of basis change in the
spinorial representation space, turns out to play a crucial role
since the sum of its eigenvalues should be equal to zero. Notice
also that for the Dirac zero modes $\Psi $, we have $\mathcal{H}\Psi
=-m\gamma _{5}\Psi $ and so a zero-mode of D with $\pm $ chirality
is also an eigenmode of $\mathcal{H}$ with eigenvalue $\lambda
\left( m\right) =\pm m$ as given here below
\begin{equation}
\mathcal{H}\Psi _{\pm }=\mp m\Psi _{\pm }.
\end{equation}%
These eigenvalues cross the axis $\lambda \left( m\right) =0$ with
two
possible slopes $\pm 1$ at $m=0$. Moreover, from the property $\mathcal{H}%
^{2}=D^{\dagger }D+m$, it follows that that $\lambda \left( m\right)
=\pm m$ are the only eigenvalues of $\mathcal{H}$ that cross the
origin at any value of $m$. Therefore the spectral flow of
$\mathcal{H}\left( m\right) $, defined as the net number of $\lambda
\left( m\right) $'s that cross the origin, counted with sign $\pm $
depending on the slope of the crossing,
comes entirely from eigenvalue crossings at $m=0$ and equals $%
N_{-}-N_{+}=-Ind\left( \mathcal{D}\right) $. Numerical results
showed that the topological index is indeed given by minus the
spectral flow hamiltonian; for more details see
\textrm{\cite{2BB,4K}.}

\subsection{Filling factor and chiral anomaly}

First recall the expression of the filling factor $\nu $ in terms of
the
number $N_{f}$ of fermions and the flux number $N_{\phi }$,%
\begin{equation}
\nu =\frac{N_{f}}{N_{\phi }}.
\end{equation}%
In 2-dimensions, we found that the filling factor $\nu
_{2D}=g_{2D}\times \left( 2N+1\right) $ where $g_{2D}$ is some
degeneracy factor giving the number of species and their quantum
numbers. The integer $\left( 2N+1\right) $ is the sum of two
contributions $\left( N+\frac{1}{2}\right) $ coming from electrons
and $\left( N+\frac{1}{2}\right) $ from holes. The two half integers
$\frac{1}{2}$ are associated with the fundamental state $N=0$ where
the conducting (electrons) and valence (holes) bands touch; this
state is a chiral zero mode as shown on eq(\ref{CZ}). Notice that
the background filed
is given by%
\begin{equation}
\mathcal{F}_{\mu \nu }^{\left( 2D\right) }=\left(
\begin{array}{cc}
0 & B \\
-B & 0%
\end{array}%
\right)
\end{equation}%
with flux $\Phi =BS$ through a surface $S$. The same situation
happens in 4-dimensions where we have found that $\nu
_{4D}=g_{4D}\times \left( 2N_{1}+1\right) \left( 2N_{2}+1\right) $.
The main difference is that in 4D
the background field involves two kinds of magnetic fields%
\begin{equation}
\mathcal{F}_{\mu \nu }^{\left( 4D\right) }=\left(
\begin{array}{cccc}
0 & B_{1} & 0 & 0 \\
-B_{1} & 0 & 0 & 0 \\
0 & 0 & 0 & B_{2} \\
0 & 0 & -B_{2} & 0%
\end{array}%
\right)
\end{equation}%
and then two fluxes $\Phi _{1}=B_{1}S$ and $\Phi _{2}=B_{2}S$. The numbers $%
\left( 2N_{1}+1\right) $ and $\left( 2N_{2}+1\right) $ are then
associated with the fluxes $\Phi _{1}$ and $\Phi _{2}$. Here also
the zero modes are chiral and so contribute to the topological
index.\newline More \ explicit relations can be written down by
working on lattices on
which the magnetic fields and the fluxes are quantized as follows,%
\begin{equation}
\begin{tabular}{l|l|l}
Lattice & magnetic fields & fluxes \\ \hline
2D & $\frac{1}{B}=\frac{L_{1}L_{2}}{2\pi N_{1}}$ & $Q_{2D}=N_{1}$ \\
&  &  \\
4D & $\frac{1}{B_{1}}=\frac{L_{1}L_{2}}{2\pi N_{1}},\quad \frac{1}{B_{2}}=%
\frac{L_{3}L_{4}}{2\pi N_{2}}$ & $Q_{4D}=2N_{1}N_{2}$ \\
&  &  \\ \hline
\end{tabular}
\label{SR}
\end{equation}%
On the other hand, using the fact that the momenta $p_{\mu }=\hbar
k_{\mu }$ of a particle of coordinate $\left( x^{\mu }\right) $ \ in
background fields
is given by the gauge covariant derivatives $iD_{\mu }=i\partial _{\mu }+%
\frac{1}{2}\mathcal{F}_{\mu \nu }x^{\nu }$; \ we learn that in the
large magnetic field this momenta (wave vector) is dominated by the
cyclotronic term $\frac{1}{2}\mathcal{F}_{\mu \nu }x^{\nu }$. This
limit leads to
\begin{equation}
\left[ x^{\mu },x^{\nu }\right] =4i\mathcal{G}^{\mu \nu }
\end{equation}%
where $\mathcal{G}^{\mu \nu }$ is as in eq(\ref{GG}). This relation
is a typical phase space relations; it teaches us that in presence
of a strong
background field the space gets discretized into fundamental cells of area $%
l_{B}^{2}=4\mathcal{G}^{\mu \nu }$. In 2-dimensions, we have, up to
a normalization factor, the following non commutative geometry
relation
\begin{equation}
\left[ x,y\right] =-\frac{i}{B}=-i\frac{L_{1}L_{2}}{2\pi N}
\end{equation}%
where the second equality, which is valid for $N\neq 0$ in agreement
with the large $B$ limit, follows from (\ref{SR}). From this
relation, we learn the area $l_{B}^{2}$ of the fundamental cell
\begin{equation}
l_{B}^{2}=\frac{L_{1}L_{2}}{2\pi N},
\end{equation}%
and so one is left with $N$ electrons (N holes) coupled to the
quantum flux leading to the filling $\nu _{2D}^{\ast }=g_{2D}\times
2N$. But this is not exactly the computed value $\nu
_{2D}=g_{2D}\times \left( 2N+1\right) $; the latter may be then
viewed as the quantum version of $\nu _{2D}^{\ast }$ and, due to the
chiral anomaly, corresponds to shifting the magnetic length as
\begin{equation}
l_{B}^{2}\rightarrow \frac{L_{1}L_{2}}{2\pi \left(
N+\frac{1}{2}\right) }, \label{SH}
\end{equation}%
which in turns corresponds to shifting the topological charge as $%
Q=N\rightarrow N+\frac{1}{2}$.\newline In 4-dimensions, the
situation is quite similar to 2D case; except that here
we have two kinds of commutations relations%
\begin{equation}
\begin{tabular}{lll}
$\left[ x,y\right] $ & $=-\frac{i}{B_{1}}=-i\frac{L_{1}L_{2}}{2\pi
N_{1}}$ &
, \\
$\left[ z,\tau \right] $ &
$=-\frac{i}{B_{2}}=-i\frac{L_{3}L_{4}}{2\pi N_{2}}
$ & ,%
\end{tabular}%
\end{equation}%
and so two kinds of magnetic lengths namely $l_{B_{1}}^{2}=\frac{L_{1}L_{2}}{%
2\pi N_{1}}$ and $l_{B_{2}}^{2}=\frac{L_{3}L_{4}}{2\pi N_{2}}$
leading afterwards to a classical relation of the filling factor
$\nu _{4D}^{\ast }=g_{4D}\times $ $4N_{1}N_{2}$. However to get the
right expression of the
filling factor namely $\nu _{4D}=g_{4D}\times $ $4\left( N_{1}+\frac{1}{2}%
\right) \left( N_{2}+\frac{1}{2}\right) $, one has to shift the
topological charges and the magnetic lengths as in (\ref{SH}); this
behavior is also a manifestation of the 4D chiral anomaly.

\section{Acknowledgements}

The authors would like to thank Drs M. Bousmina, M. Daoud and A.
Jellal for discussions. L.B Drissi thanks the associateship program
of ICTP, Trieste, Italy. E.H Saidi thanks URAC 09/CNRST.

\end{document}